\documentclass[twocolumn,superscriptaddress,floatfix,preprintnumbers]{revtex4-1}
\usepackage{graphics,amssymb,amsmath,epsfig,color}
\usepackage{graphicx}
\usepackage{subfigure}
\begin{document}

\title{Tuneable quantum interference in a 3D integrated circuit}

\author{Zachary Chaboyer}
\affiliation{Centre for Ultrahigh bandwidth Devices for Optical Systems (CUDOS), MQ Photonics Research Centre, Department of Physics and Astronomy, Macquarie University, NSW 2109, Australia}
\email{zachary.chaboyer@students.mq.edu.au}
\author{Thomas Meany}
\affiliation{Centre for Ultrahigh bandwidth Devices for Optical Systems (CUDOS), MQ Photonics Research Centre, Department of Physics and Astronomy, Macquarie University, NSW 2109, Australia}
\author{L. G. Helt}
\affiliation{Centre for Ultrahigh bandwidth Devices for Optical Systems (CUDOS), MQ Photonics Research Centre, Department of Physics and Astronomy, Macquarie University, NSW 2109, Australia}
\author{Michael J. Withford}
\affiliation{Centre for Ultrahigh bandwidth Devices for Optical Systems (CUDOS), MQ Photonics Research Centre, Department of Physics and Astronomy, Macquarie University, NSW 2109, Australia}
\author{M. J. Steel}
\affiliation{Centre for Ultrahigh bandwidth Devices for Optical Systems (CUDOS), MQ Photonics Research Centre, Department of Physics and Astronomy, Macquarie University, NSW 2109, Australia}

\begin{abstract}
Integrated photonics promises solutions to questions of stability, complexity, and size in quantum optics. Advances in tunable and non-planar integrated platforms, such laser-inscribed photonics, continue to bring the realisation of quantum advantages in computation and metrology ever closer, perhaps most easily seen in multi-path interferometry. Here we demonstrate control of two-photon interference in a chip-scale 3D multi-path interferometer, showing a reduced periodicity and enhanced visibility compared to single photon measurements. Observed non-classical visibilities are widely tunable, and explained well by theoretical predictions based on classical measurements.  With these predictions we extract a Fisher information approaching a theoretical maximum, demonstrating the capability of the device for quantum enhanced phase measurements.
\end{abstract}
\maketitle

When an $N$ photon Fock state passes through an optical delay it behaves as a single
object with a momentum $N$ times larger than that of the corresponding single
photon state~\cite{Kok2010}.  For path-entangled photons in the two arms of a Mach-Zehnder
interferometer, either $N$ in the upper arm and 0 in the lower, or 0 in the
upper arm and $N$ in the lower (so called ``N00N'' states), this phenomenon
leads to a reduced peak to peak distance in interference fringes seen at the
output, enabling quantum-enhanced phase estimation~\cite{Xiang2011,Mitchell2004,Nagata2007} and, in theory, higher-resolution lithography~\cite{Boto2001}. More recently,
attention has been devoted to further improving the precision of quantum
interferometric schemes by employing optimized input
states~\cite{Dorner2009,Xiang2013} and adaptive feedback
phases~\cite{Lovett2013,Wheatley2010}, as well as developing schemes that are
robust to photon loss~\cite{Matthews2013}.  However, the majority of these
efforts have employed conventional two-path Mach-Zehnder and Michelson
interferometers. 

Additional gains can be realised  by increasing the number of paths available
for photons to take in an interferometer~\cite{Greenberger2000}, including protocols for multiparameter
estimation~\cite{Spagnolo2012,Humphreys2013} and a reduced sensitivity to
photon loss~\cite{Cooper2011}. However, limitations of existing technology have
to this point prevented experimental demonstrations of non-classical
interference in multi-path interferometers. Although such devices can be
implemented relatively straightforwardly using single-mode fibre
components~\cite{Weihs1996}, utilizing these fibre devices for quantum
interferometry is made difficult by the phase instability caused by thermal and
acoustic noise~\cite{Weihs1996a}. For this reason, experimental quantum
interferometry in multi-arm devices has remained largely unexplored despite its
potential advantages.

Here we report on the quantum enhancement in fringe periodicity observed when passing pairs of
photons through two integrated three-port splitters placed in series
to form a three-path analogue of a Mach-Zehnder
interferometer~\cite{D'Ariano1997} (see Fig.~\ref{fig:Setup}).  Multiphoton interference at the first
three-way splitter, or ``tritter'', causes the photons to coalesce into a
superposition of photons occupying each interferometer
arm~\cite{Campos2000,Meany2012,Spagnolo2013}. The photons then probabalistically pass
through the same phase in the measurement arm, resulting in a known subset of two-photon fringes exhibiting both a reduced periodicity and higher visibility than would be obtained classically. By mounting a thermo-optic phase shifter onto the chip, we are able to adjust the relative phase in one of the interferometer arms. This tunability allows us to control the visibility of the two-photon quantum interference that occurs within the device, thereby allowing for a measurement of two-photon fringes with enhanced periodicity. We close with a discussion on the applicability of our device to quantum-enhanced phase measurement.

\section{Results}
\subsection{Device fabrication and illumination}
\begin{figure}[htbp] \centering
{\label{fig:1}\includegraphics[width=0.45\textwidth]{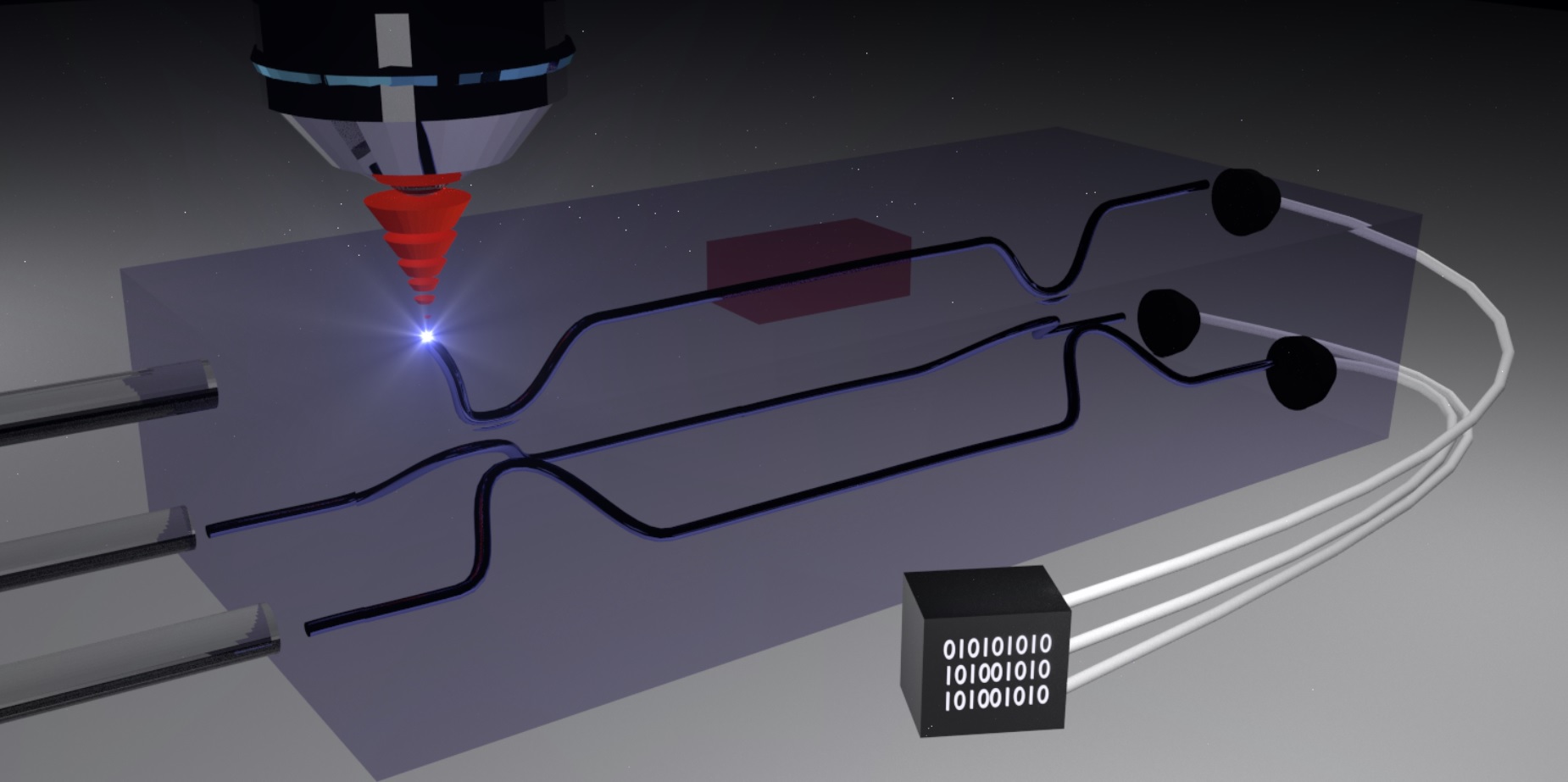}}
\caption{ 
Design of the 3D interferometer, the direct laser inscription method employed here enables the single step fabrication of a truly 3D circuit. The middle interferometer arm is raised with respect the others, imparting a relative phase due to interaction with the chip surface.
  \label{fig:Setup}
  }
\end{figure}

While precise control of non-classical interference has been demonstrated in
two-arm Mach-Zehnder interferometers fabricated using planar
silica-on-silicon~\cite{Matthews2009a} and UV-writing
technologies~\cite{Smith2009}, these platforms require multiports to be
implemented as multimode interference devices~\cite{Peruzzo2011}, which have
low fabrication tolerances, or as complex concatenations of 50/50 beamsplitters
and phase shifters~\cite{Metcalf2012}. Therefore, we employed the femtosecond
laser direct-write (FLDW) technique~\cite{Davis1996} to enable stable multi-arm quantum
interferometry in a 40~mm long alumino-borosilicate glass chip (see Methods). The middle interferometer arm 
was raised 127 $\mu$m above the other two arms to allow for interaction with a heating element mounted on the chip surface that acts as a thermo-optic phase
shifter (see Fig.~\ref{fig:Setup}).

The fabricated three-arm interferometer was excited using photons from
a type I spontaneous parametric down-conversion (SPDC) source that produced
degenerate photons at 804~nm. The photons were coupled in and out of the
chip using optical fibres aligned in V-groove arrays.  Single photon detection 
was performed using silicon avalanche photodiodes (see Methods for details).

\subsection{Circuit characteristics}

In characterising the device,
the majority of losses occur outside the chip in
the form of coupling losses at the facets and detector efficiencies, while
low propagation losses are achieved in the waveguide
section~\cite{Meany2013,Arriola2013}. 
The interferometer may thus be well approximated mathematically by an operation $\hat{U}(\theta)$. Here $\theta$ is the relative phase applied to the middle interferometer arm which we control using a heating element mounted on the glass surface (see Fig.~\ref{fig:Setup}) via the thermo-optic effect. A state $\left| \psi_\text{in}\right\rangle $ injected into the device evolves into the state $\left| \psi_\text{out}\right\rangle = \hat{U}(\theta) \left| \psi_\text{in}\right\rangle $, and in the photon spatial mode basis representation the unitary operator takes the form of a $3\times 3$ matrix connecting the input modes $a_i^\text{in}$ to the output modes $a_i^\text{out}$ (see supplementary information).
\begin{equation}
(a_i^\text{out})^\dagger=\sum_j U_{ij}(\theta) (a_j^\text{in})^\dagger.
\end{equation}

\begin{figure}[htbp]
  \centering
 % \subfigure[]{\label{fig:2}\includegraphics[width=0.4\textwidth]{./figures/Singles1jvsVoltage.jpg}}
 % \subfigure[]{\label{fig:3}\includegraphics[width=0.4\textwidth]{./figures/U1j.jpg}}
 \subfigure[]{\includegraphics[width=0.45\textwidth]{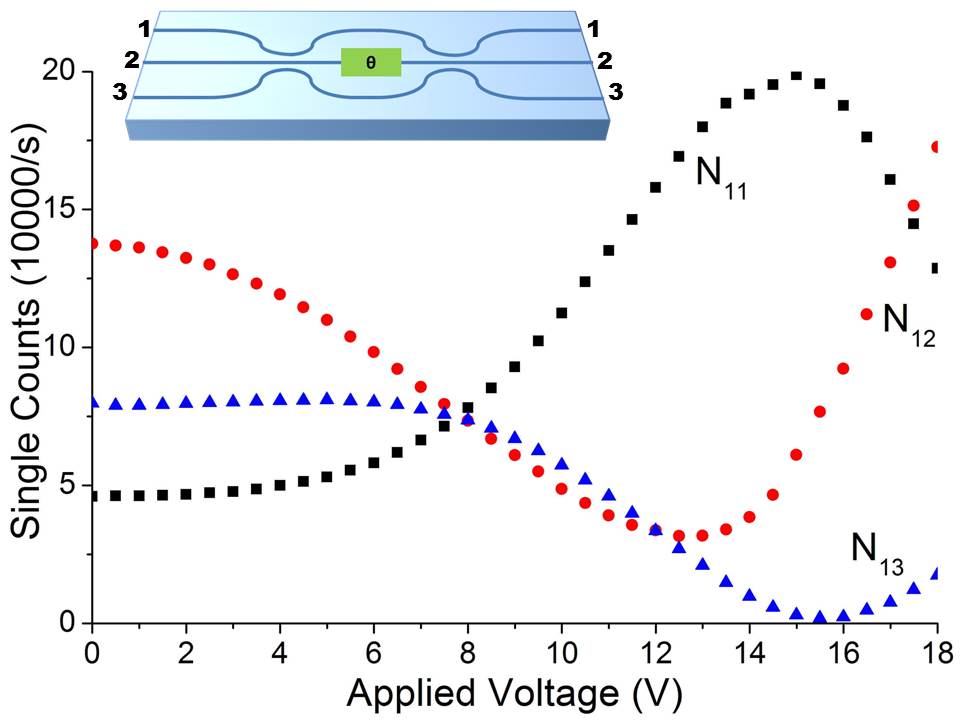}}
 \subfigure[]{\includegraphics[width=0.45\textwidth]{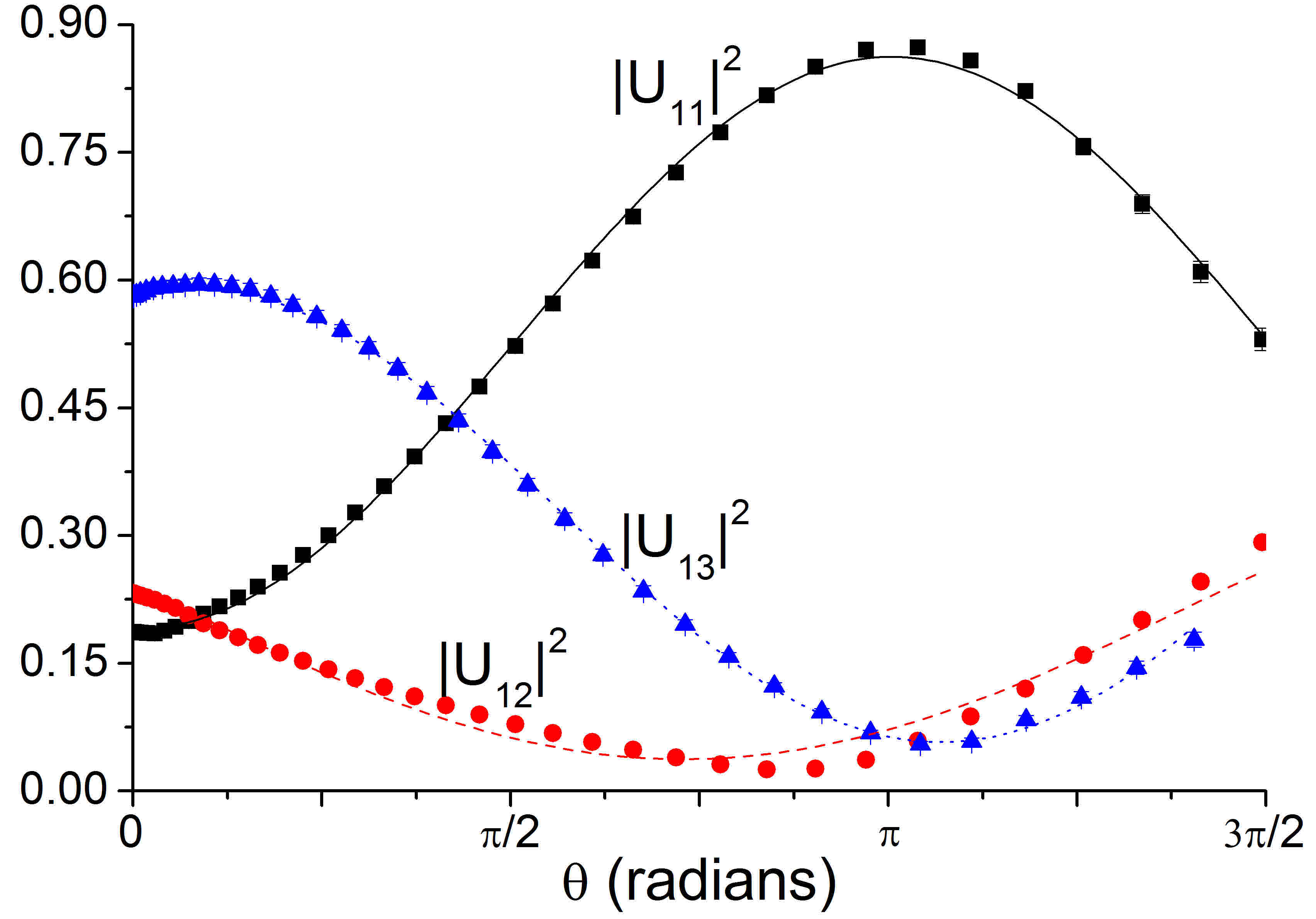}}
 \vspace{-0.4cm}
%  \subfigure[]{\label{fig:5}\includegraphics[width=0.3\textwidth]{./figures/PredictedTwoPhoton}}
  \caption{(a) Single photon fringes measured when launching into port 1. Poisson error bars are smaller than the data points and are omitted. Inset: schematic of the device showing the numbering convention for the input and output ports. (b) $|U_{1j}|^{2}$ extracted from the raw single photon data plotted as a function of induced phase. Black squares: $|U_{11}|^{2}$, red circles: $|U_{12}|^{2}$, blue triangles: $|U_{13}|^{2}$, curves: fits to $|U_{1j}(\theta)|^{2}=A\sin{\theta}+B\cos{\theta}$.
  \label{fig:fringes}
  }
\end{figure}

The nonlinear relationship between the induced phase $\theta$ and the applied
voltage $V$ was established by single-photon characterisation of the nine possible 
classical interference patterns, examining the count rates from each pair of output ports
for separate excitation of each input port. 
Performing single-photon measurements using the same SPDC source allowed us
to ensure consistency between our classical characterisation and the subsequent non-classical study. 
The measured counts
as a function of voltage when injecting into input port~1 are shown in
Fig~\ref{fig:fringes}(a). 
Since the input and output coupling losses
at each port $ \eta_i^\text{in}, \eta_j^\text{out}$ are unknown,
the unitary matrix elements $U_{ij}(\theta(V))$ at each voltage were determined by 
maximum-likelihood estimation applied to ratios of the measured
count rates that are independent of the facet losses~\cite{Meany2012} (see supplementary material).
To first approximation, the thermo-optic phase change depends linearly on the
dissipated power as $\theta=k IV$, where $I$ is the current in the heater at
given voltage $V$. The matrix elements $U_{ij}(\theta)$ contain phase terms of the form $e^{i\theta}$, and their squared moduli were fitted to a function taking the form
$|U_{ij}(k IV)|^{2}=A\sin{(kIV)}+B\cos{(kIV)}$. 
After normalisation and
fitting, we determine a proportionality constant of $k=0.626\pm 0.031 \mbox{
W}^{-1}$. The
normalised and fitted classical fringes for input 1 are shown in
Fig.~\ref{fig:fringes}(b) (see supplementary material for others).

\subsection{Quantum Characterisation} 
\begin{figure}[htbp]
  \centering
  %\subfigure[]{\label{fig:12}\includegraphics[width=0.3\textwidth]{./figures/Scans1212.pdf}}
  %\subfigure[]{\label{fig:13}\includegraphics[width=0.3\textwidth]{./figures/Scans1223.pdf}}
  %\subfigure[]{\label{fig:4}\includegraphics[width=0.4\textwidth]{./figures/Scans1213.jpg}}
  %\subfigure[]{\label{fig:5}\includegraphics[width=0.4\textwidth]{./figures/V12mn.jpg}}
  \subfigure[]{\includegraphics[width=0.45\textwidth]{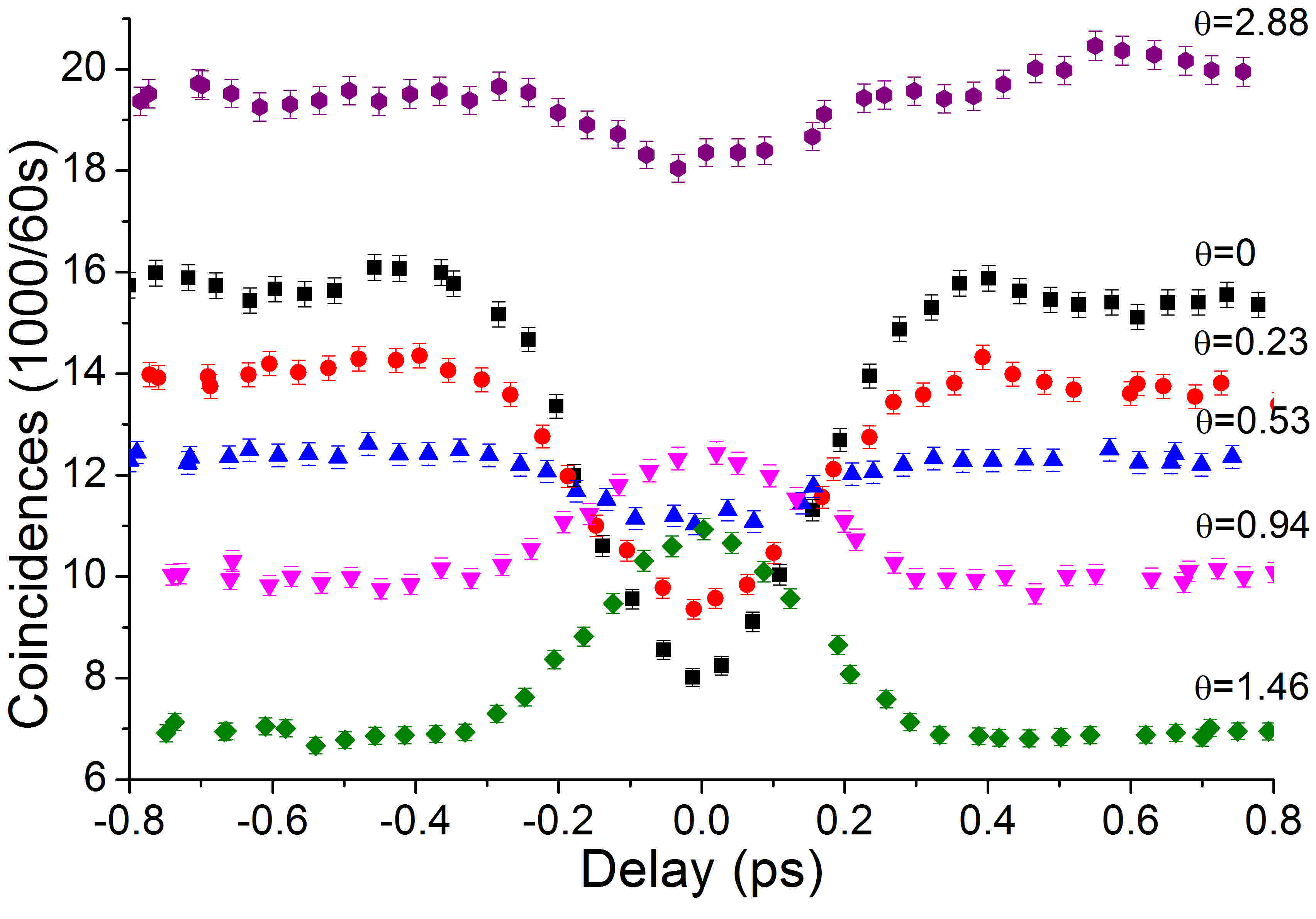}}
  \subfigure[]{\includegraphics[width=0.45\textwidth]{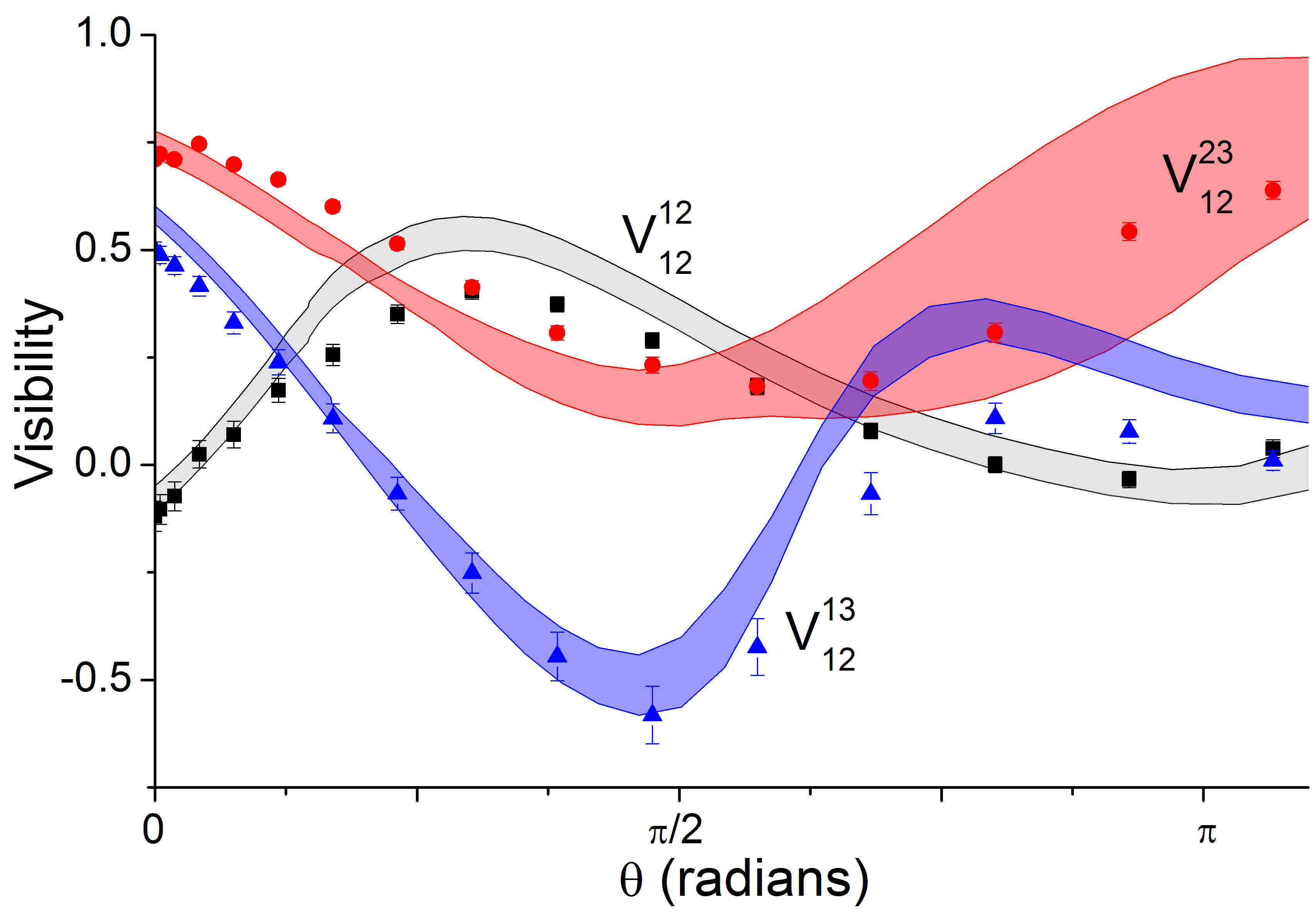}}
  \vspace{-.4cm}
\caption{(a) Two-photon coincidences as a function of relative delay at various values of phase $\theta$ injecting into inputs 1 and 2 and measuring at outputs 1 and 3. (b) Measured (points) and predicted (bands) two-photon visibilities as a function of $\theta$ when injecting $\left| 110\right\rangle$. Black squares: $V_{12}^{12}$, red circles: $V_{12}^{23}$, blue triangles: $V_{12}^{13}$. Grey band: predicted upper and lower bounds for $V_{12}^{12}$, red band: predicted bounds for $V_{12}^{23}$, blue band: predicted bounds for $V_{12}^{13}$.
\label{fig:TwoPhoton}
}
\end{figure}

We then proceeded to the quantum
characterisation of the tunable device by injecting photon pairs into each
combination of input ports, corresponding to the Fock states $\left|
\psi_\text{in}\right\rangle = \left| 110 \right\rangle, \left| 011 \right\rangle,
\left| 101 \right\rangle $. The distinguishability of the photons was
controlled by adjusting their relative arrival time at the chip by means of a
free space delay. Scans of the delay $\tau$ were performed while varying
$\theta$ and measuring coincidence counts between photons emerging from each
combination of output ports. The change in the output state $\left|
\psi_\text{out}\right\rangle$ with $\theta$ was monitored as a change in the degree
of quantum interference in a two-photon experiment as quantified by the
visibility
\begin{equation} \label{eq:Visibility}
V_{ij}^{mn}(\theta) = \frac{C_{ij}^{mn}(\tau_\text{max})-C_{ij}^{mn}(\tau=0)}{C_{ij}^{mn}(\tau_\text{max})}.
\end{equation}
Here $C_{ij}^{mn}(\tau_\text{max})$ is the measured coincidences at outputs $m$
and $n$ when injecting into inputs $i$ and $j$ when the delay is maximized and
$C_{ij}^{mn}(\tau=0)$ is the measured counts at minimum distinguishability
as estimated from a Gaussian fit. The coincidence probability at minimal
distinguishability depends on the relative phases of the elements $U_{ij}(\theta)$. This leads to both
constructive interference in which a coincidence peak is observed or
destructive interference, seen as a coincidence dip, as the relative phase
$\theta$ is tuned by the thermo-optic element. This effect is shown in
Fig.~\ref{fig:TwoPhoton}(a), in which we see a coincidence dip at $\theta=0$~rad 
become a coincidence peak at $\theta=0.94$~rad. The measured visibilities
after subtracting accidental coincidence counts for each pair of output ports
when injecting $\left| 110 \right\rangle$ are shown in
Fig.~\ref{fig:TwoPhoton}(b). These are compared with the visibilities predicted from the classical characterisation (shaded curves, see supplementary material). Here we see agreement between the theory and experiment within error over most of the range of $\theta$.

\begin{figure}
  \centering
  \includegraphics[width=0.45\textwidth]{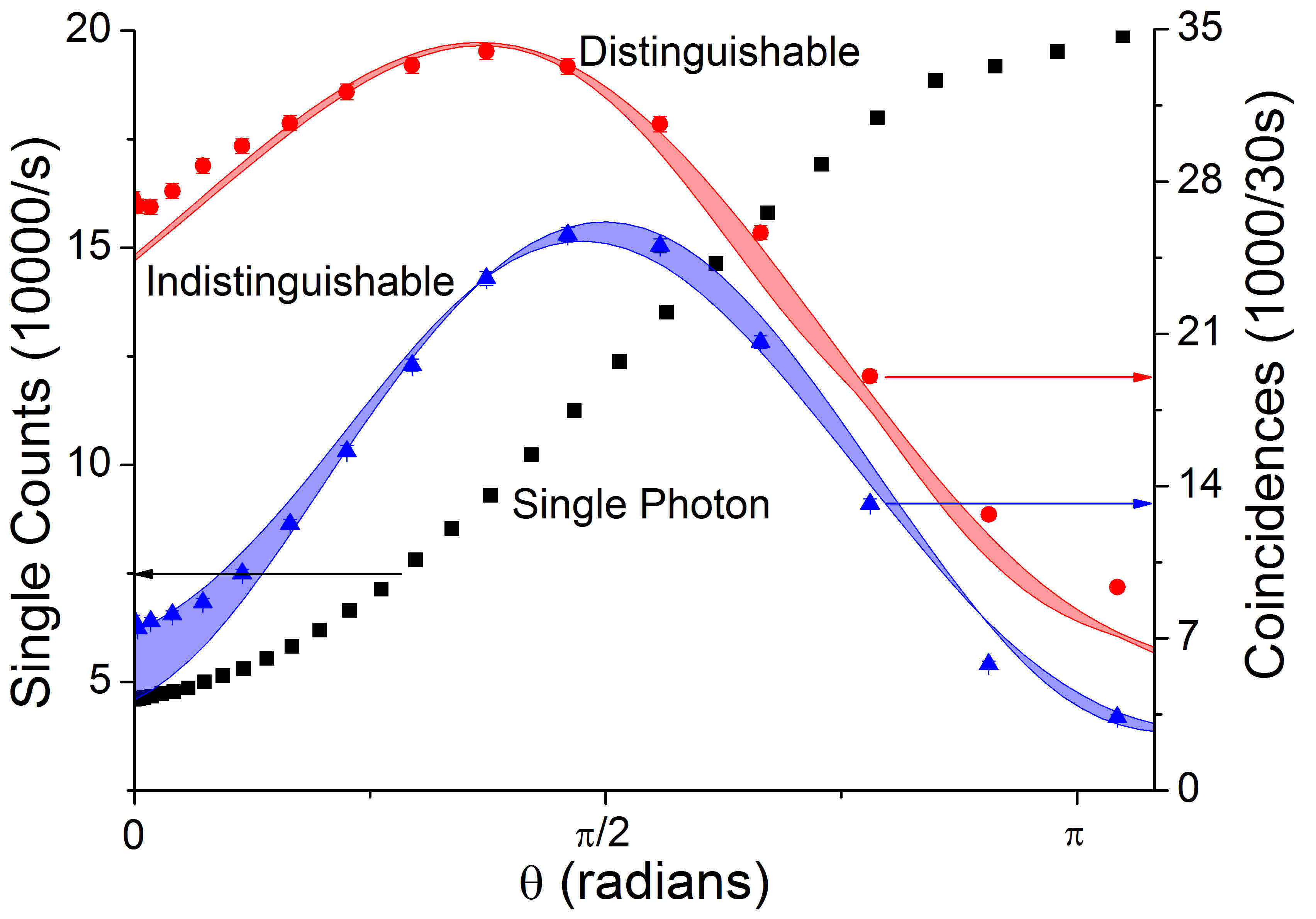}
%  \subfigure[]{\includegraphics[width=0.3\textwidth]{./figures/Q23mn_measured.pdf}}
%  \subfigure[]{\includegraphics[width=0.3\textwidth]{./figures/Q13mn_measured.pdf}}
  \vspace{-.38cm}
\caption{Measured single and coincidence counts as a function of phase. Black squares and left axis: single counts injecting into port 1 and measuring at port 1. Red and right axis: measured (circles) and predicted (band) coincidence counts at outputs 2 and 3 when injecting into ports 1 and 2 at maximal distinguishability. Blue and right axis: measured (triangles) and predicted (band) coincidence counts at outputs 2 and 3 when injecting into ports 1 and 2 at minimal distinguishability.
\label{fig:ExptQs}
}
\end{figure}

Two-photon non-classical fringes are obtained from the data described above by taking the coincidence counts at the point of minimal distinguishability. The resulting coincidence counts at outputs 2 and 3 when injecting into ports 1 and 2 are plotted in Fig.~\ref{fig:ExptQs}~(blue triangles) along with the theoretical prediction $C_{12}^{23}=\eta_{1}^{\text{in}}\eta_{3}^{\text{out}}\eta_{2}^{\text{in}}\eta_{2}^{\text{out}}p_{12}^{23}(\tau=0)C_{\text{in}}$. Here the relevant pairs of losses $\eta_{i}^{\text{in}}\eta_{j}^{\text{out}}$ were solved for from the single photon measurements using the previously extracted unitary. These are overlayed with the single counts $N_{11}$ (black squares) as well as the coincidences in the distinguishable case (red) for the same input and output ports, exhibiting a reduction in periodicity of both two-photon fringes compared with the single photon case. This is expected as there is a probability of two photons passing through the same interferometer arm in both the distinguishable and indistinguishable cases. However, the visibility is increased from $57 \%$ classically to $77 \%$ in the non-classical case due to the suppression of coincidences by quantum interference.

\section{Discussion}

\begin{figure}[htbp]
  \centering
  {\label{fig:7}\includegraphics[width=0.45\textwidth]{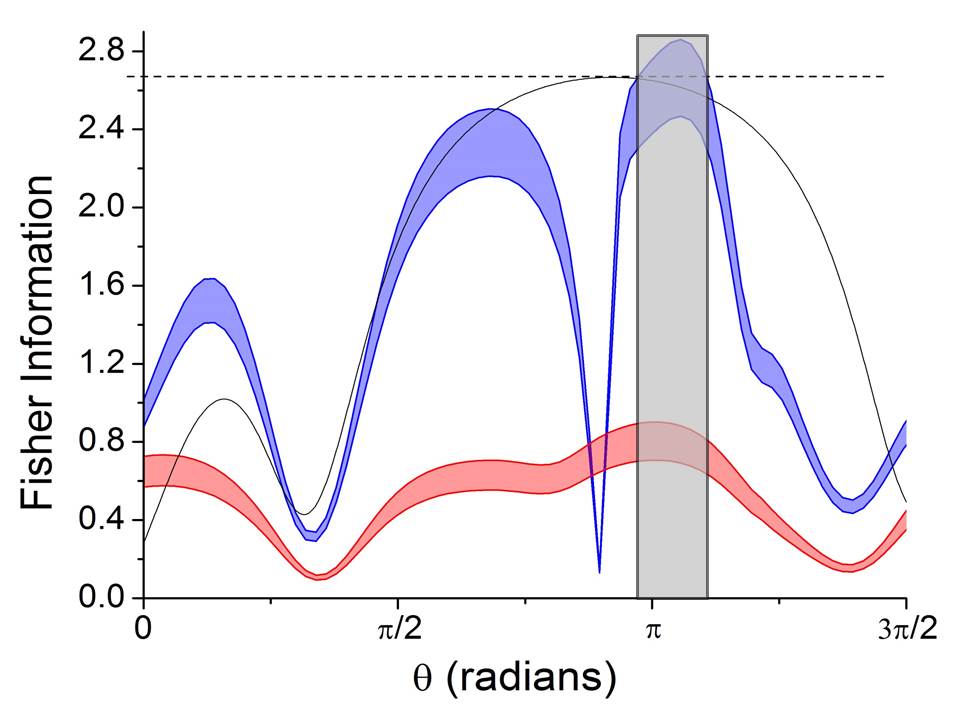}}
  \vspace{-.4cm}
\caption{Extracted Fisher information of the three-arm interferometer when injecting $\vert 011\rangle$ (blue curve) and single photons (red curve). Grey shaded region: values of the phase $\theta$ where the theoretical limit for an ideal three-path interferometer lies within the tolerance limit of the extracted Fisher information. The Fisher information calculated for an ideal, symmetric device with a two-photon input state is plotted (black curve) for comparison.
\label{fig:fisher}
}
\end{figure}

The analysis performed to this point now allows us to quantify the potential of our device for quantum metrology by extracting the classical Fisher information possible with our interferometer. The Fisher information $F\left(\theta\right)$ is a measure of the amount of information that can be gained about an unknown measurand $\theta$ by sampling the measurement outcomes of a given probe system~\cite{Kok2010}. In this case, the measurand is the phase $\theta$ and the measurement outcomes are the number of photons present in each output mode. The Fisher information is then calculated by summing over all possible sets of nonclassical interference fringes $p_{ijk} = |\left\langle ijk \right| \hat{U}(\theta)\left| \psi_\text{in}\right\rangle|^{2}$ as follows:
\begin{equation} \label{eq:Fisher}
F\left( \theta \right) =\sum_{i,j,k=0}^2 \dfrac{1}{p_{ijk}}\left( \dfrac{\partial p_{ijk}}{\partial \theta}\right) ^{2}.
\end{equation}
The Fisher information obtained from Eq.~\eqref{eq:Fisher} with the $\vert 011\rangle$ input state is shown in Fig.~\ref{fig:fisher} (blue), surpassing the values obtainable with single photon inputs (red). The theoretical maximum of $8/3 \approx 2.67$ obtainable for photon pairs injected into an ideal three-arm interferometer also lies within the tolerance of the calculated $F(\theta)$ between $0.97\pi$ and $1.10\pi$ (grey shaded region). The sharp decrease seen at $\theta \approx 0.89\pi$ is due to the fringes $p_{110}$, $p_{011}$, $p_{020}$ and $p_{002}$ all reaching an extremum at approximately the same point. Since this subset of fringes make the dominant contribution and the single fringe Fisher information for each is dependent on its first derivative, the result is a minimum in the total $F(\theta)$. A Fisher information of $F>N$, where $N$ is the number of photons in the probe state, is required to realize a quantum enhancement in an interferometric phase measurement~\cite{Matthews2013}. Therefore, our result shows that phase measurements below the standard limit should be possible with our device.

The variation of the Fisher information with the phase $\theta$ leads to a measurement precision that depends on the unknown phase. This, however may be counteracted by making use of an adaptive measurement scheme in which the device is tuned back to the value of $\theta$ corresponding to the optimal $F(\theta)$ using an adjustable feedback phase~\cite{Spagnolo2012}. Such a scheme can be straightforwardly implemented in our device by using the demonstrated thermo-optic phase shifter as feedback, while the unknown phase would be imparted by another element such as a microfluidic channel~\cite{Crespi2012a}.

It is important to note that the phase $\theta$ in this device remained stable over the full duration of the experiment and that the measured visibilities were reproduced within the given errors despite the modest effort taken to stabilize it against temperature fluctuations and vibration. The stability and reproducibility of the phase supports the utility of our device for the adaptive measurement schemes described above. This is contrasted with equivalent fibre devices that have been reported to date, in which thermal fluctuations of the optical path of several wavelengths limited the study to a characterisation with a classical laser source~\cite{Weihs1996,Weihs1996a}. Placing separate phase shifters on another one of the interferometer arms would allow for the simultaneous measurement of two phases~\cite{Spagnolo2012,Humphreys2013}. This is a new and mostly unexplored type of measurement that is unique to multi-arm devices such as the one reported here. Furthermore this device could be used as a novel multiphoton state generator through integration with on chip single photon sources~\cite{Meany2014,Pryde2003}.

We have reported the first experimental demonstration of quantum interferometry in a 3D tunable laser-written three-port interferometer.  The integrated geometry provided by FLDW enables improved stability compared to fibre and bulk set-ups, and the tunability illustrates the enhanced sensitivity achievable using non-classical states. The combination of high stability, robust 3D fabrication and tunability makes this device an ideal platform for practical and highly sensitive quantum sensing and an enabling technology for new multi-parameter estimation experiments.

\section*{Methods}
\noindent\textbf{Device fabrication}

The interferometer was fabricated in alumino-borosilicate glass (Corning
Eagle2000) by femtosecond laser direct-writing. This technique uses
a femtosecond laser, focused inside a transparent substrate, to form localised
refractive index change. The substrate can be translated with respect to the
focus, thereby forming lines of index change which can act as
waveguides~\cite{Davis1996}. The laser can be focused at multiple depths inside
the sample meaning truly 3D waveguide circuits can be
formed~\cite{Crespi2013a}, with high transmission~\cite{Meany2013,Arriola2013}. The output
of a Ti:Sapphire oscillator (Femtolasers GmbH, FEMTOSOURCE XL 500, 800 nm
centre wavelength, 5.1 MHz repetition rate, $<$50 fs pulse duration) was
focussed into the sample using a 100$\times$ oil immersion objective. The
sample was translated accross the beam focus using Aerotech motion control
stages with 10 nm precision. A pulse energy of 31~nJ and a sample translation
speed of 2000~mm/min were selected in order to yield waveguides with single
mode operation at 800 nm. The waveguides were ellipsoidal in shape and
supported modes with a diameter of approximately 6.5~$\mu$m in the semi-major
axis and 4.6~$\mu$m in the semi-minor axis. The waveguides follow a raised sine
curve from an initial spacing of 127~$\mu$m to a 7~$\mu$m spacing in the 1.1~mm
long interaction region over a distance of 8.175~mm. The four raised sine curves, two interaction regions and a straight section on the interferometer arms give a total device length of 40~mm. \\

\noindent\textbf{Photon pair generation}

The fabricated three-arm interferometer was characterised using photons from
a spontaneous parametric down-conversion (SPDC) source. The output of a
continuous-wave, 402 nm laser diode (Toptica iBeam smart) was focussed into a 1
mm thick type-I phase matched BiBO crystal. Down converted photons were
produced at 804 nm with a $6^{\circ}$ opening cone angle and were passed through 3
nm bandpass filters before being coupled into polarization maintaining single
mode fibres. One of the fibres was mounted on a computer controlled micron
resolution actuator which translated one collection fibre with respect to
another leading to a relative path delay. This was used to tune the temporal
distinguishibility of the photons. The fibres were then butt-coupled to the
waveguides at the chip facets using commercial fibre V-groove arrays with a
fibre spacing of 127 $\mu$m matching the waveguide spacing at the facets of the
fabricated device. Photons were coupled out of the chip at the opposite facet
by another V-groove array with standard single-mode fibres before being
monitored by a set of silicon avalanche photodiodes (Excelitas).

\section*{Funding Information}
This research was supported by the ARC Centre of Excellence for Ultrahigh bandwidth Devices for Optical Systems (project number CE110001018),
and performed in part at the Optofab node of the Australian National Fabrication  Facility; a company established under the National Collaborative Research  Infrastructure Strategy to  provide nano and microfabrication facilities for Australian  researchers.

\section*{Acknowledgments}
The authors thank Zhizhong Yan for technical support.

%%%%%%%%%% Merge with supplemental materials %%%%%%%%%%
\widetext
\clearpage
\begin{center}
\textbf{\large Tunable quantum interference in a 3D integrated circuit\\
-Supplementary Material-}
\end{center}
%%%%%%%%%% Merge with supplemental materials %%%%%%%%%%
%%%%%%%%%% Prefix a "S" to all equations, figures, tables and reset the counter %%%%%%%%%%
\setcounter{equation}{0}
\setcounter{figure}{0}
\setcounter{table}{0}
\setcounter{page}{1}
\makeatletter
\renewcommand{\theequation}{S\arabic{equation}}
\renewcommand{\thefigure}{S\arabic{figure}}
\renewcommand{\bibnumfmt}[1]{[S#1]}
\renewcommand{\citenumfont}[1]{S#1}
%%%%%%%%%% Prefix a "S" to all equations, figures, tables and reset the counter %%%%%%%%%%

\section{Extraction of phase-dependent unitary}
\begin{figure*}[htbp]
  \centering
  \subfigure[]{\label{fig:1}\includegraphics[width=0.3\textwidth]{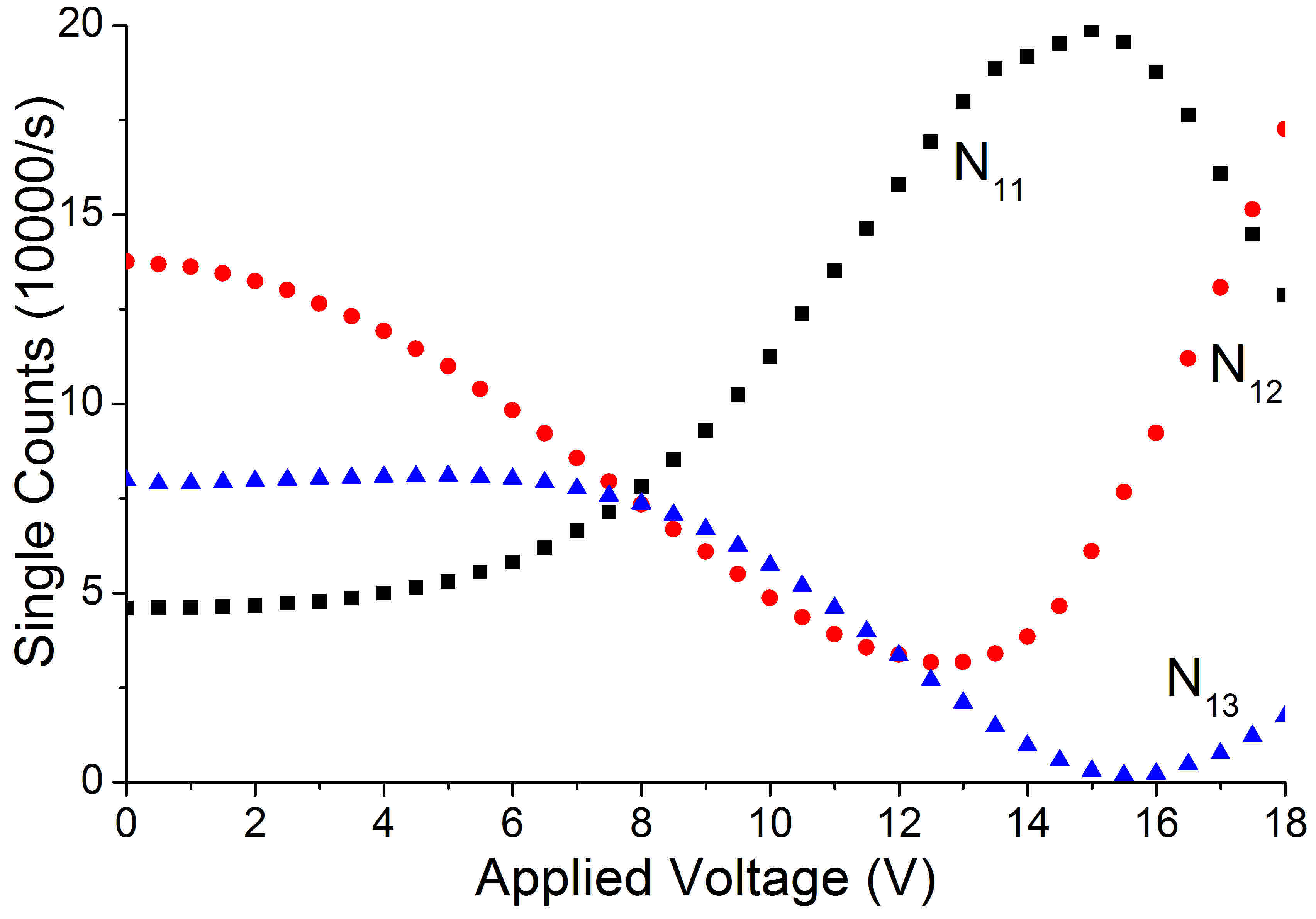}}
  \subfigure[]{\label{fig:2}\includegraphics[width=0.3\textwidth]{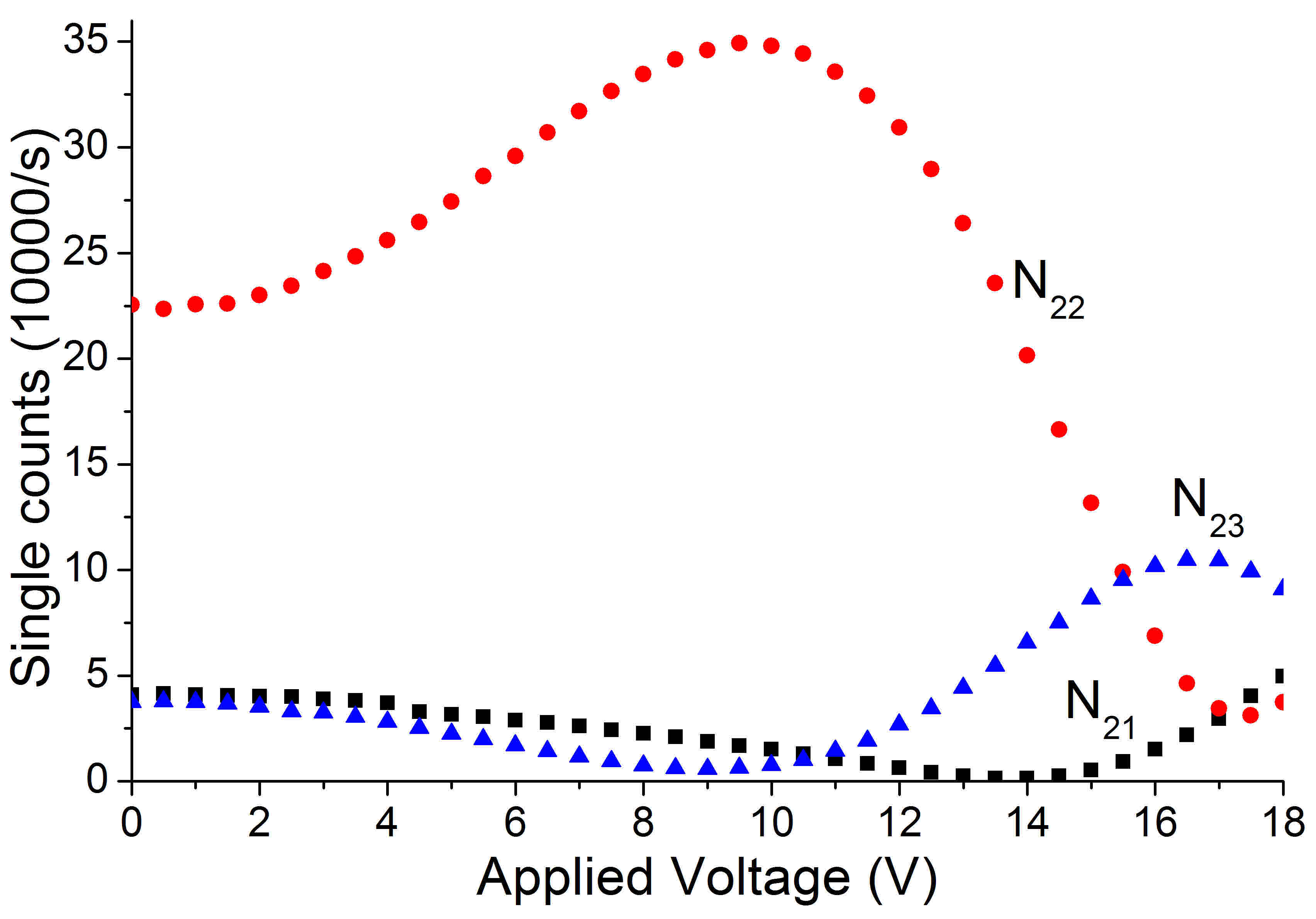}}
  \subfigure[]{\label{fig:3}\includegraphics[width=0.3\textwidth]{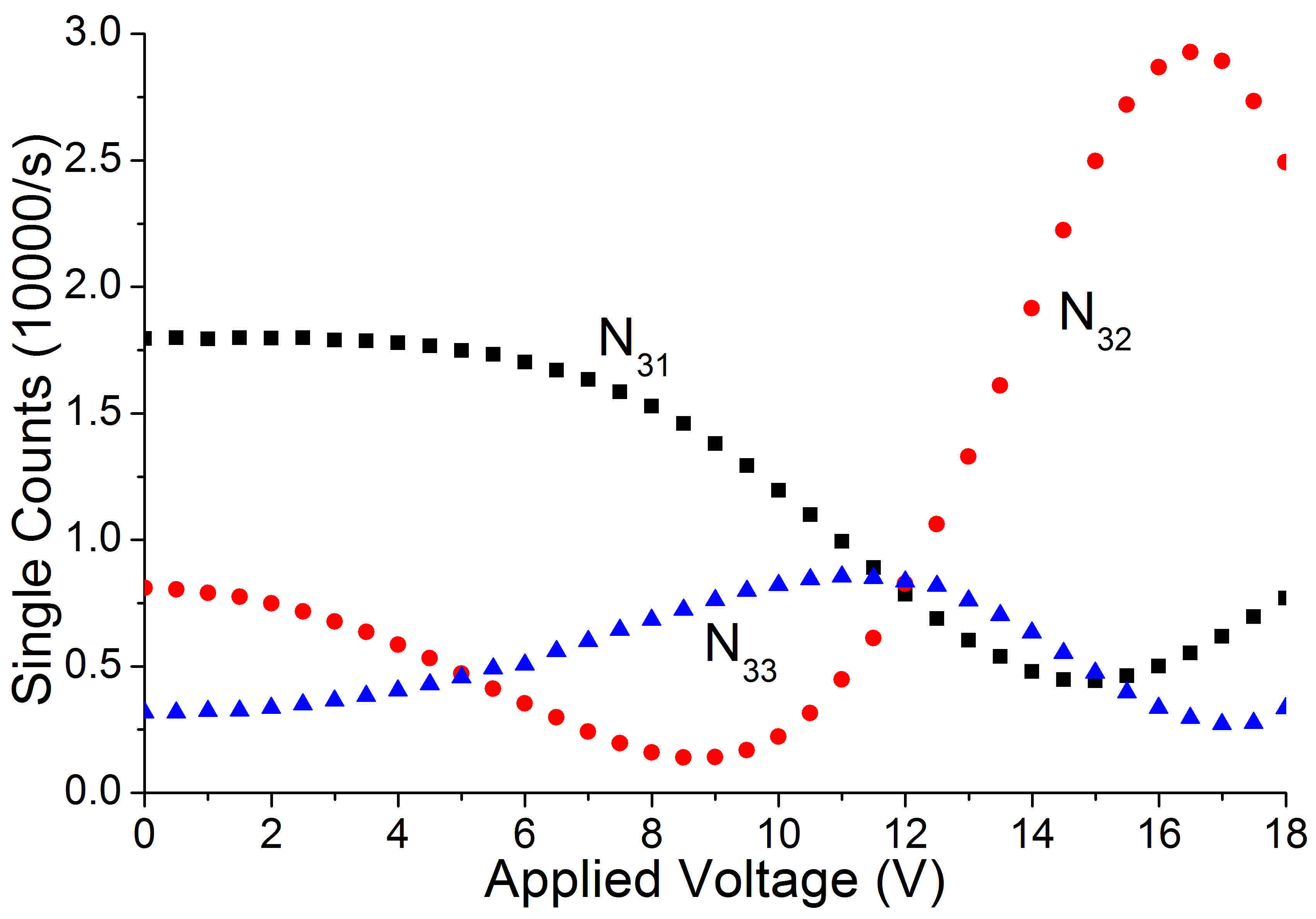}}
%  \vspace{-.4cm}
\caption{Counts at each output port as a function of voltage applied to the resistive heater when injecting single photons into (a) input 1, (b) input 2, (c) input 3. The error bars are smaller than the data points and are omitted.
\label{fig:SinglesvsVoltage}
}
\end{figure*}

Our three-arm interferometer may be represented by a $3\times 3$ unitary transformation exhibiting a dependence on the relative phase induced in the middle arm by the thermo-optic phase shifter. Here we detail the extraction of this unitary as a function of the induced phase from a set of measured classical interference fringes. For this characterisation, we use photons from a single collection arm of a photon pair source based on spontaneous parametric downconversion in a nonlinear crystal. Since classical interference is essentially a single photon phenomenon, the fringes obtained in this way are equivalent to those that would be obtained using classical bright light. Photons were injected into each input port, while single counts were monitored at each output using silicon avalanche photodiodes. The three sets of single photon counts as a function of the voltage applied to the resistive heater on the chip surface are shown in Fig.~\ref{fig:SinglesvsVoltage}.

\begin{figure*}[htbp]
  \centering
  \subfigure[]{\label{fig:4}\includegraphics[width=0.4\textwidth]{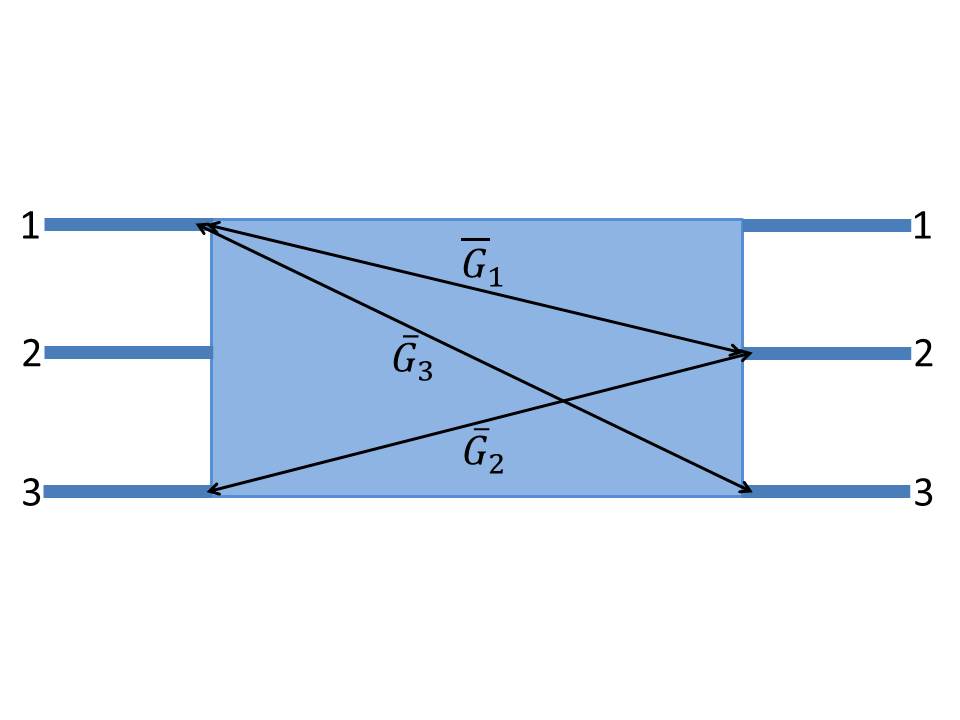}}
  \subfigure[]{\label{fig:5}\includegraphics[width=0.4\textwidth]{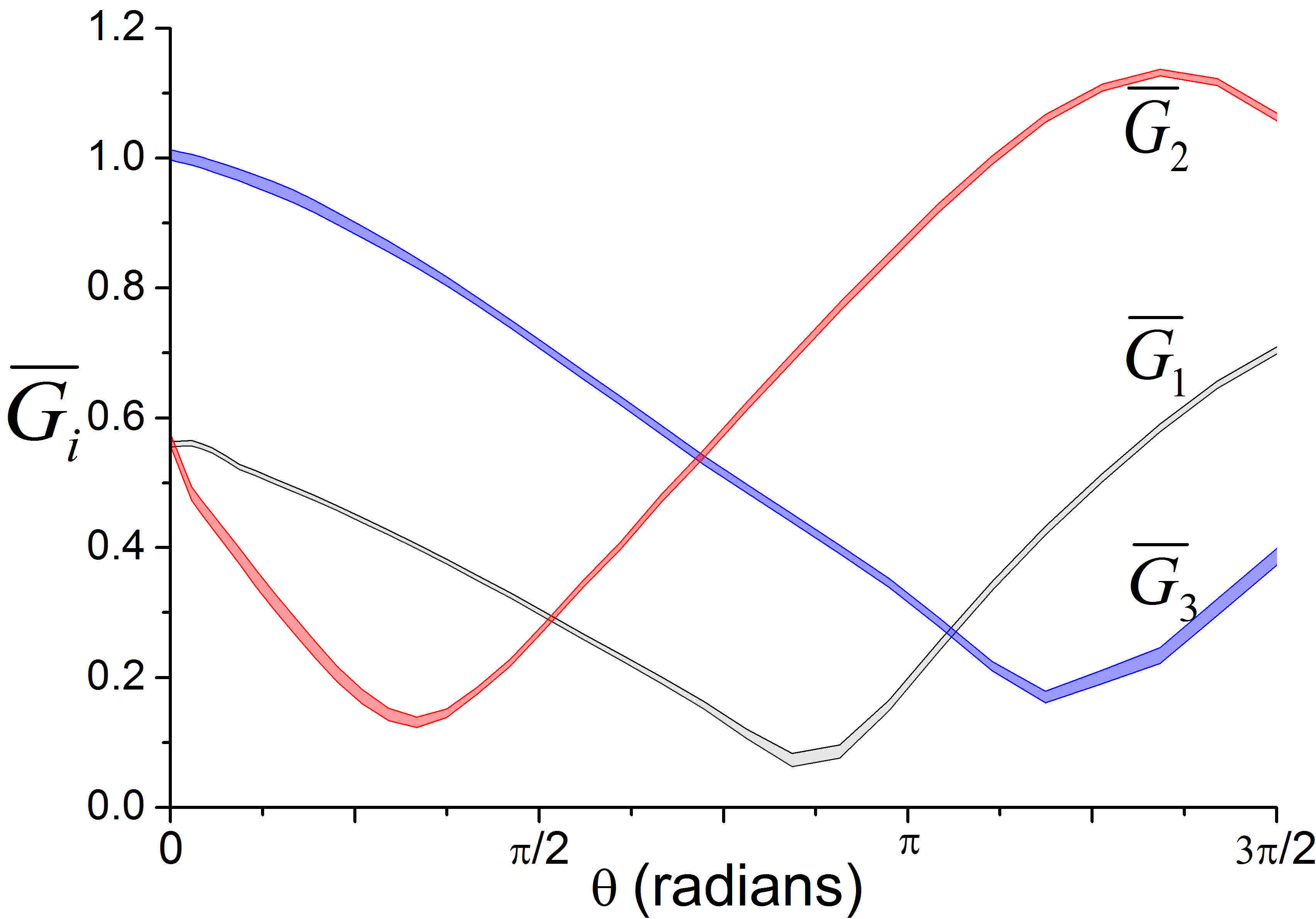}}
%  \vspace{-.4cm}
\caption{(a) Schematic of the three-port device described by effective couplings $G_{1}$, $G_{2}$ and $G_{3}$. (b) Tolerance limits (represented by shaded bands) of the extracted coupling coefficients plotted as a function of induced phase in the middle interferometer arm.
\label{fig:CouplingCoeffs}
}
\end{figure*}

\begin{figure*}[htbp]
  \centering
  \subfigure[]{\label{fig:6}\includegraphics[width=0.3\textwidth]{./figures/U1j.jpg}}
  \subfigure[]{\label{fig:7}\includegraphics[width=0.3\textwidth]{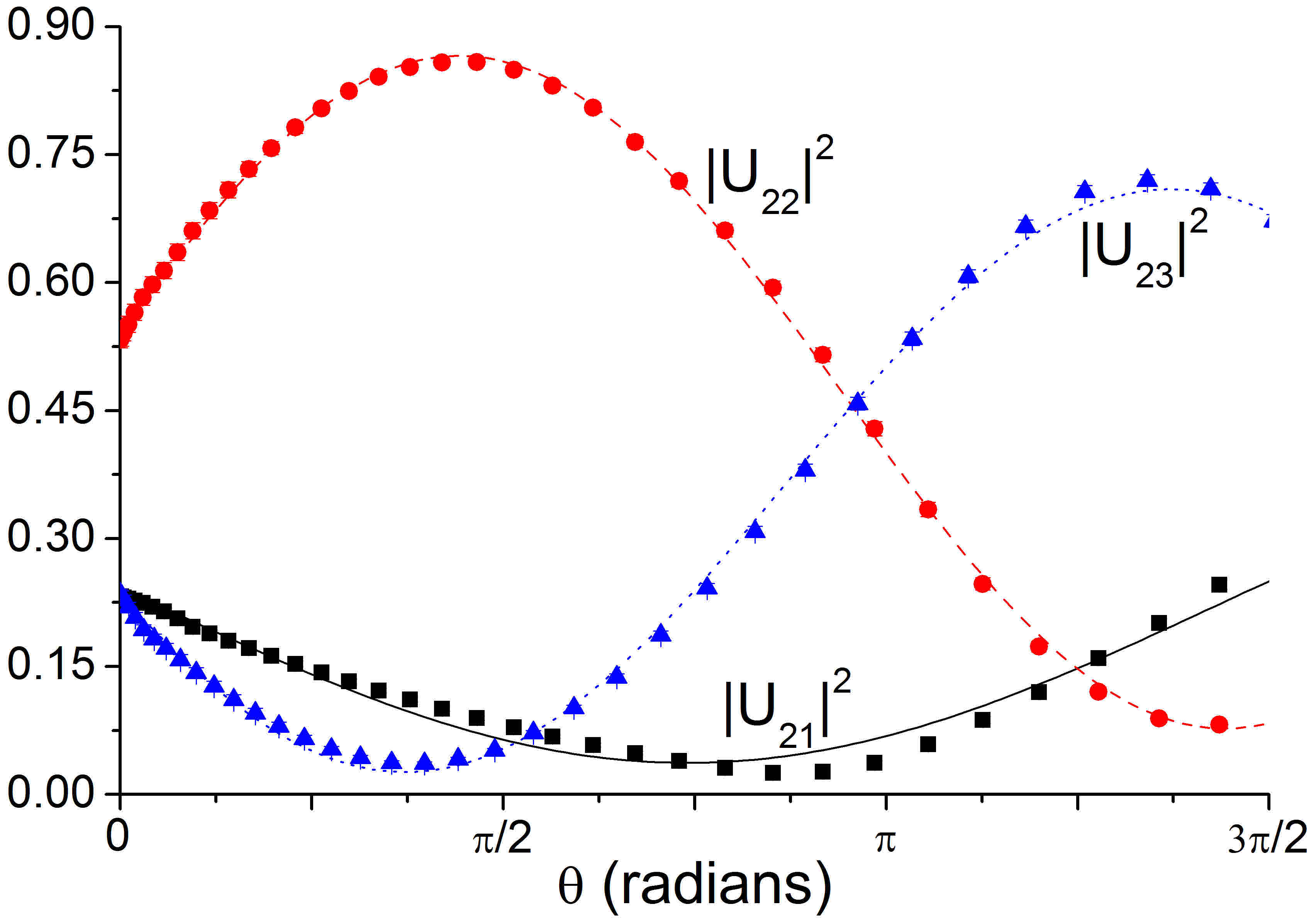}}
  \subfigure[]{\label{fig:8}\includegraphics[width=0.3\textwidth]{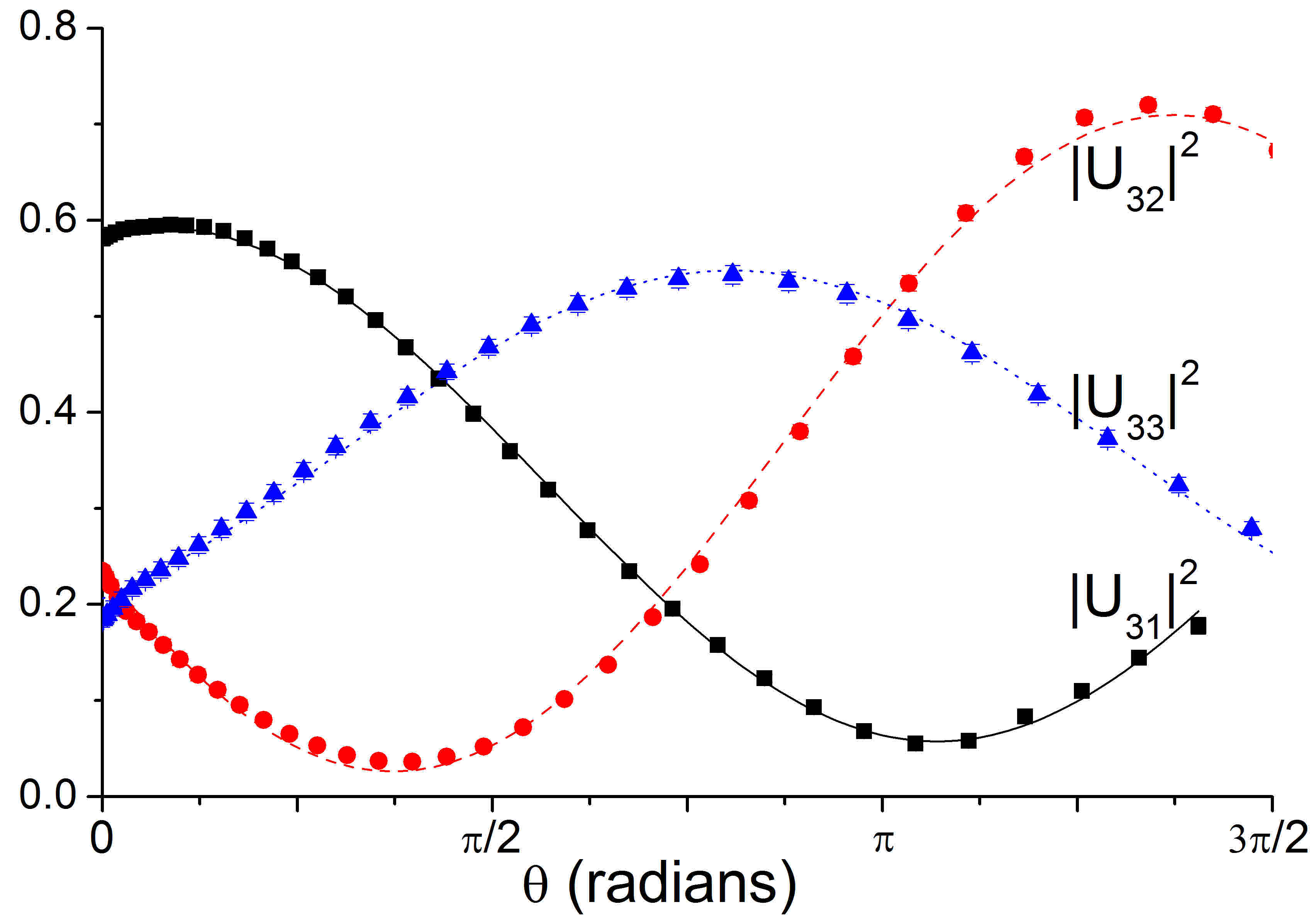}}
%  \vspace{-.4cm}
\caption{Extracted $|U_{ij}|^{2}$ as a function of induced phase for (a) $i=1$, (b) $i=2$, (c) $i=3$.
\label{fig:Umods}
}
\end{figure*}

The unitary matrix describing the device consists of both real transition amplitudes and complex phases~\cite{Laing2012}. The method described in~\cite{Laing2012} allows the transition amplitudes of an arbitrary $N$ port device to be determined from the $N^2$ possible intensity (or single photon count) measurements, while it is possible to solve for the phases using measured two-photon visibilities. However, if we wish to predict two-photon data using only single-photon measurements, we must reduce the number of free parameters. This is done by considering our interferometer as a three-port device characterised by a coupling matrix $C$ that varies with the induced phase $\theta$ in the middle arm. The coupling matrix describes the evolution of the input modes $\boldsymbol{\hat{a}}=(\hat{a}_{1},\hat{a}_{2},\hat{a}_{3})$ to the output modes $\boldsymbol{\hat{b}}=(\hat{b}_{1},\hat{b}_{2},\hat{b}_{3})$ according to
\begin{equation}
\label{eq:modeevolution}
\dfrac{d\hat{b}}{dz}=-iC(z,\theta)\hat{a},
\end{equation}
where
\begin{equation} \label{eq:couplingmatrix}
C(z,\theta)=\begin{pmatrix}
\beta & G_{1}(z,\theta) & G_{3}(z,\theta) \\
G_{1}(z,\theta) & \beta & G_{2}(z,\theta) \\
G_{3}(z,\theta) & G_{2}(z,\theta) & \beta \\
\end{pmatrix}.
\end{equation}
We solve Eq.~\eqref{eq:modeevolution} approximately by considering an effective coupling matrix $\tilde{C}(\theta)$ taking the same form as~\eqref{eq:couplingmatrix}, with effective coupling coefficients $\bar{G_{i}}$ (see Fig.~\ref{fig:CouplingCoeffs}(a)). This gives the approximate solution $\hat{b}=e^{-i\tilde{C}}\hat{a}$, allowing the unitary transformation $U(\theta)$ governing the evolution of the device's input to output modes to be determined as the matrix exponential of the effective coupling matrix. The counts $N_{ij}$ at each output port $j$ when injecting $M$ number of photons into port $i$ are related to this transfomation by
\begin{equation} \label{Single_Counts}
N_{ij}=\eta_{i}^\text{in}\eta_{j}^\text{out}\left| U_{ij}\right|^{2}M
\end{equation}
The input and output losses $\eta_{i}^\text{in}$ and $\eta_{j}^\text{out}$ (in which the output loss includes both the facet loss and the non-unit efficiency of the detector monitoring output $j$) are accounted for by forming appropriate ratios of single counts for which the losses cancel~\cite{Laing2012}
\begin{equation}
F_{ijkl}=\dfrac{N_{ij}N_{kl}}{N_{il}N_{kj}}
\end{equation}
We now fit the theoretically determined ratios $F_{1122}$, $F_{1133}$ and $F_{2233}$ to their measured values at each value of $\theta$ using a maximum likelihood technique~\cite{Meany2012}, assuming that the ratios follow a Gaussian distribution. The log-likelihood function then takes the following form weighted against the experimental errors $\sigma_{ijkl}$
\begin{equation}
L(V)=\sum \dfrac{(F_{ijkl}-\frac{\left| U_{ij}\right|^{2}\left| U_{kl}\right|^{2}}{\left| U_{il}\right|^{2}\left| U_{kj}\right|^{2}})^{2}}{2(\sigma_{ijkl})^{2}}.
\end{equation}
The function $L$ takes the form of a Gaussian near the converged values of the coupling coefficients when plotted as a function of each $G_{i}$ while holding the others constant. This allows tolerances for each to be determined based on the $1/e^{2}$ width of each distribution. The resulting upper and lower bounds for each coupling coefficient $G_{i}$ are plotted in Fig.~\ref{fig:CouplingCoeffs}(b) as shaded bands.
A value of $U(V)$ is determined at each data point using a routine that iterates through a set of nine arrays of single count measurements, minimizing $L(V)$ at each data point using a simplex algorithm. This yields a set of coupling coefficients as a function of the electrical power $P=IV$ dissipated by the heater. We can then determine the induced phase by fitting the power-dependent data to the expression $|U_{ij}(\theta)|^{2}=A\sin{(kIV)}+B\cos{(kIV)}$. The obtained $|U_{ij}|^{2}$ are plotted in Fig.~\ref{fig:Umods} and are equivalent to a normalisation of the measured fringes accounting for the different facet losses at each input and output port.

\section{Quantum Characterisation}

\begin{figure*}[htbp]
  \centering
  \subfigure[]{\label{fig:12}\includegraphics[width=0.3\textwidth]{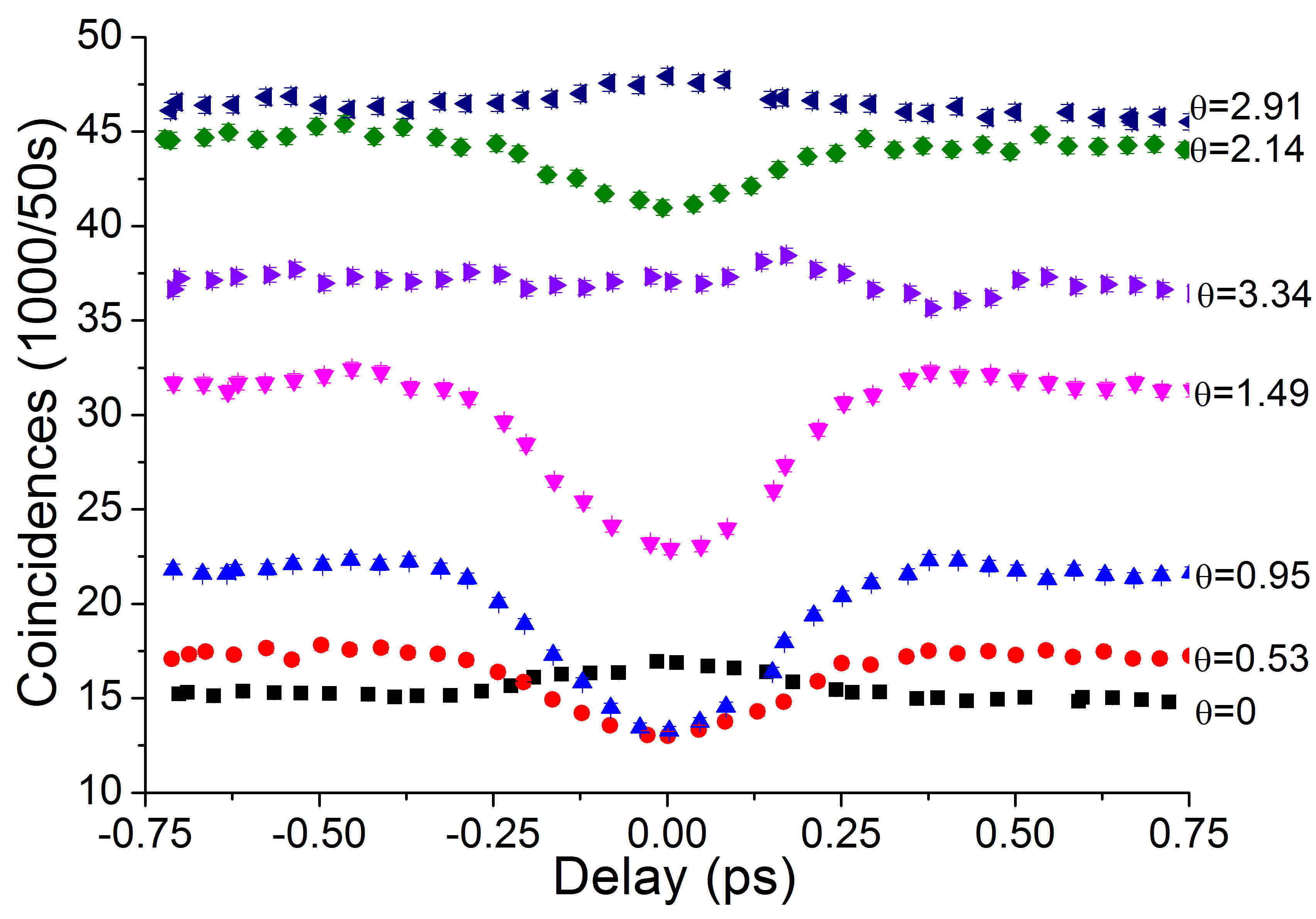}}
  \subfigure[]{\label{fig:13}\includegraphics[width=0.3\textwidth]{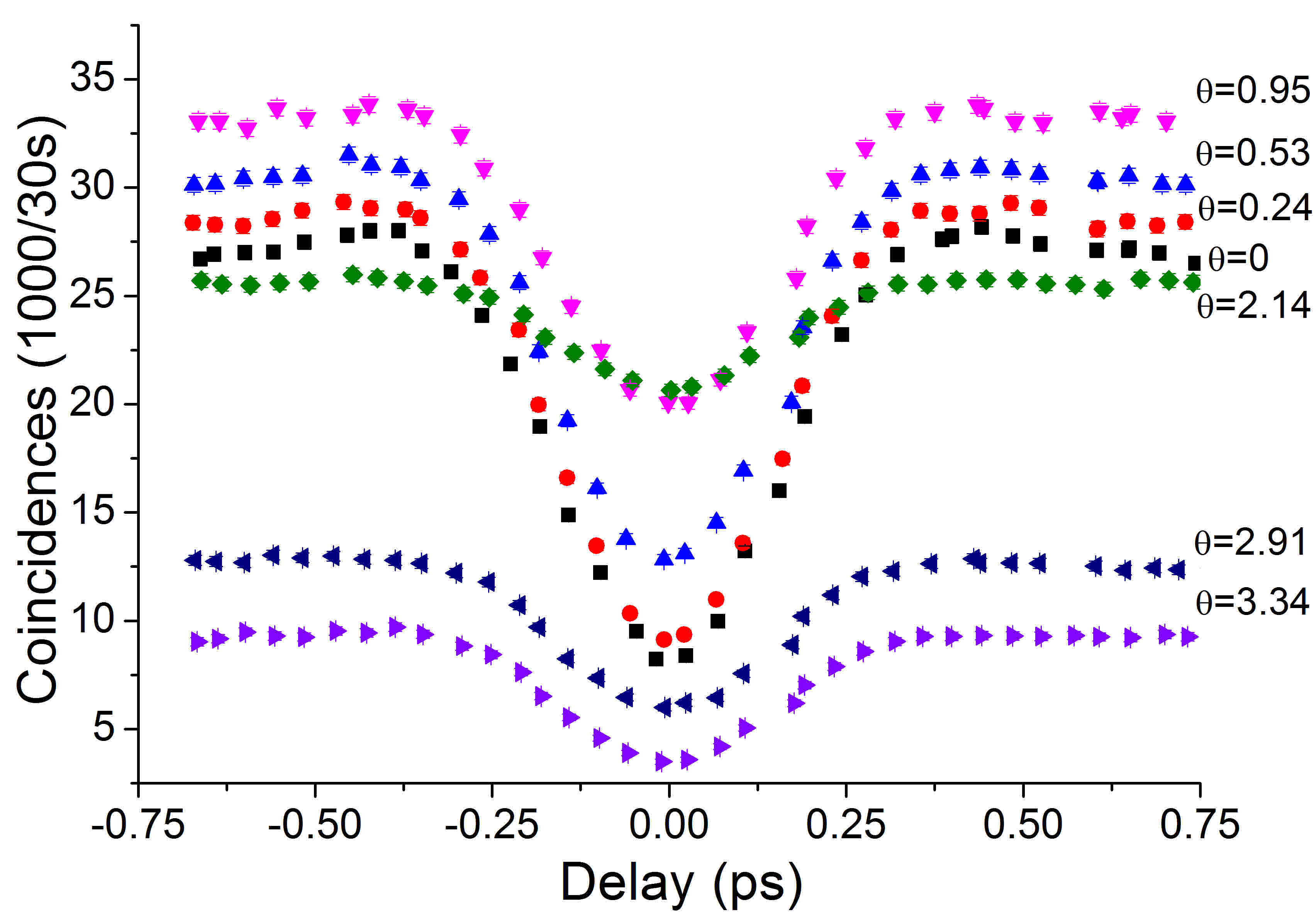}}
  \subfigure[]{\label{fig:14}\includegraphics[width=0.3\textwidth]{./figures/Scans1213.jpg}}
%  \vspace{-.4cm}
\caption{Two-photon coincidences as a function of relative delay at various values of heater voltage while injecting $\left| 110\right\rangle$ and (a) measuring $\left| 110\right\rangle$, (b) measuring $\left| 011\right\rangle$, (c) measuring $\left| 101\right\rangle$.
\label{fig:HOMDipsin12}
}
\end{figure*}

\begin{figure*}[htbp]
  \centering
  \subfigure[]{\label{fig:15}\includegraphics[width=0.3\textwidth]{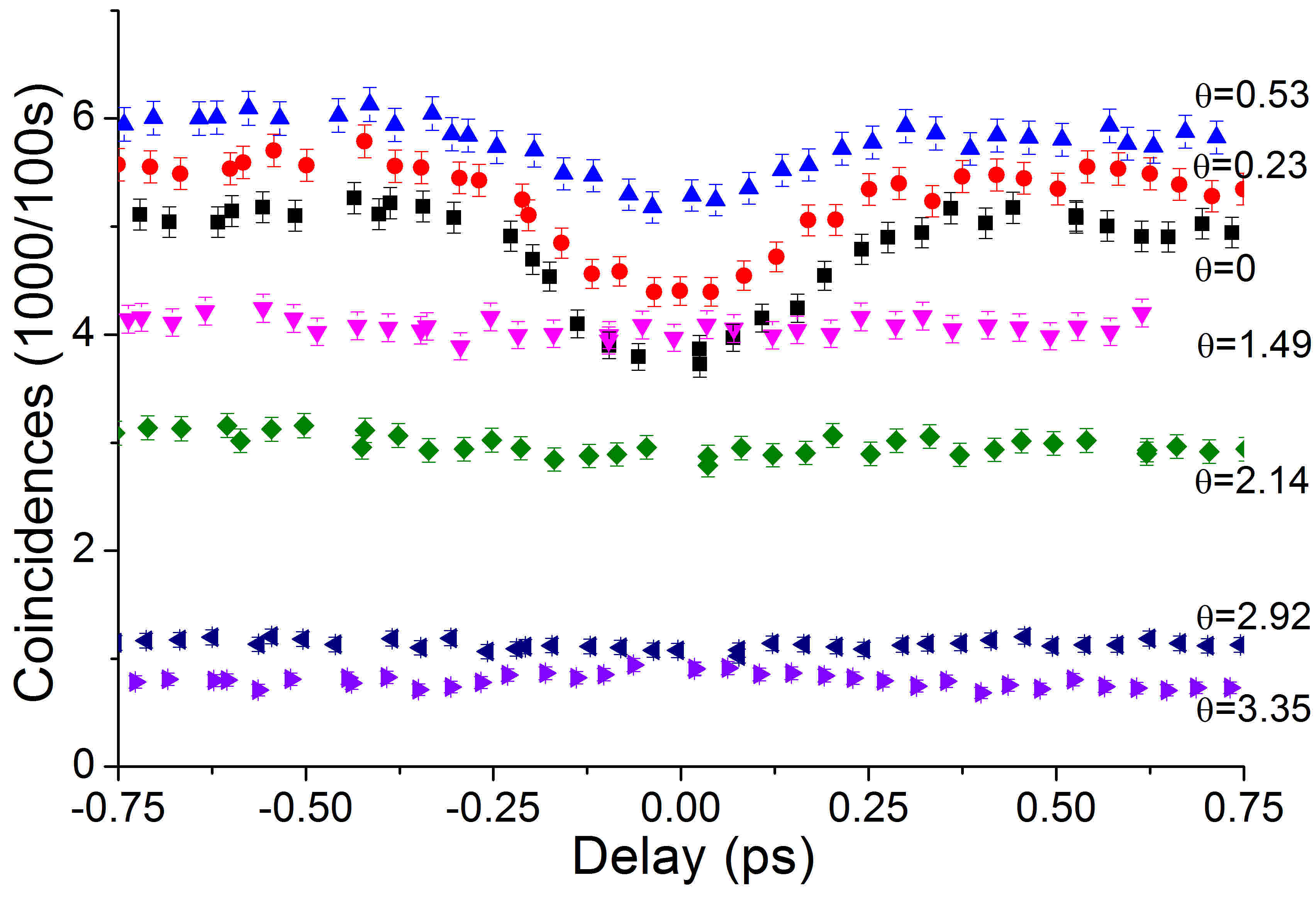}}
  \subfigure[]{\label{fig:16}\includegraphics[width=0.3\textwidth]{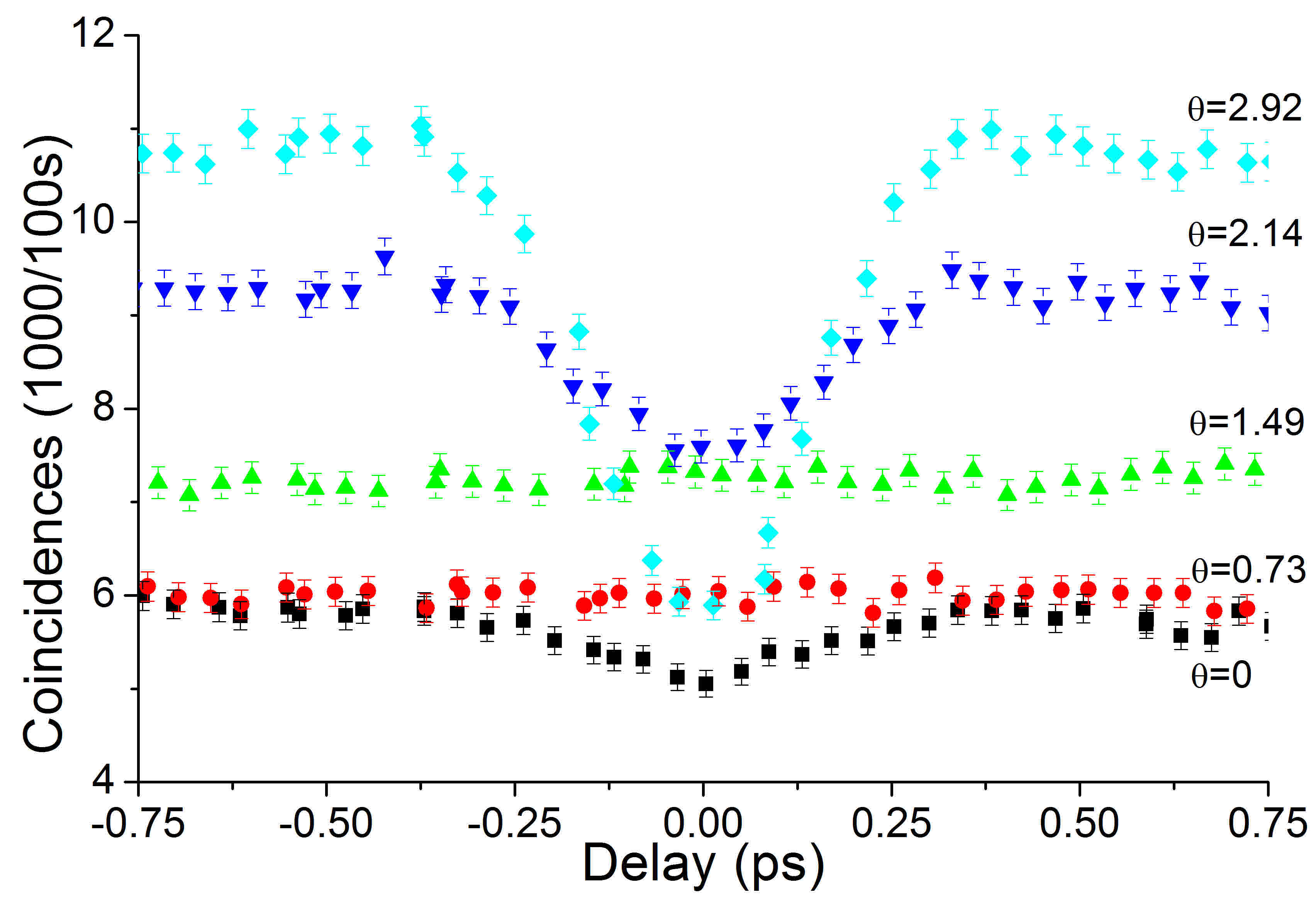}}
  \subfigure[]{\label{fig:17}\includegraphics[width=0.3\textwidth]{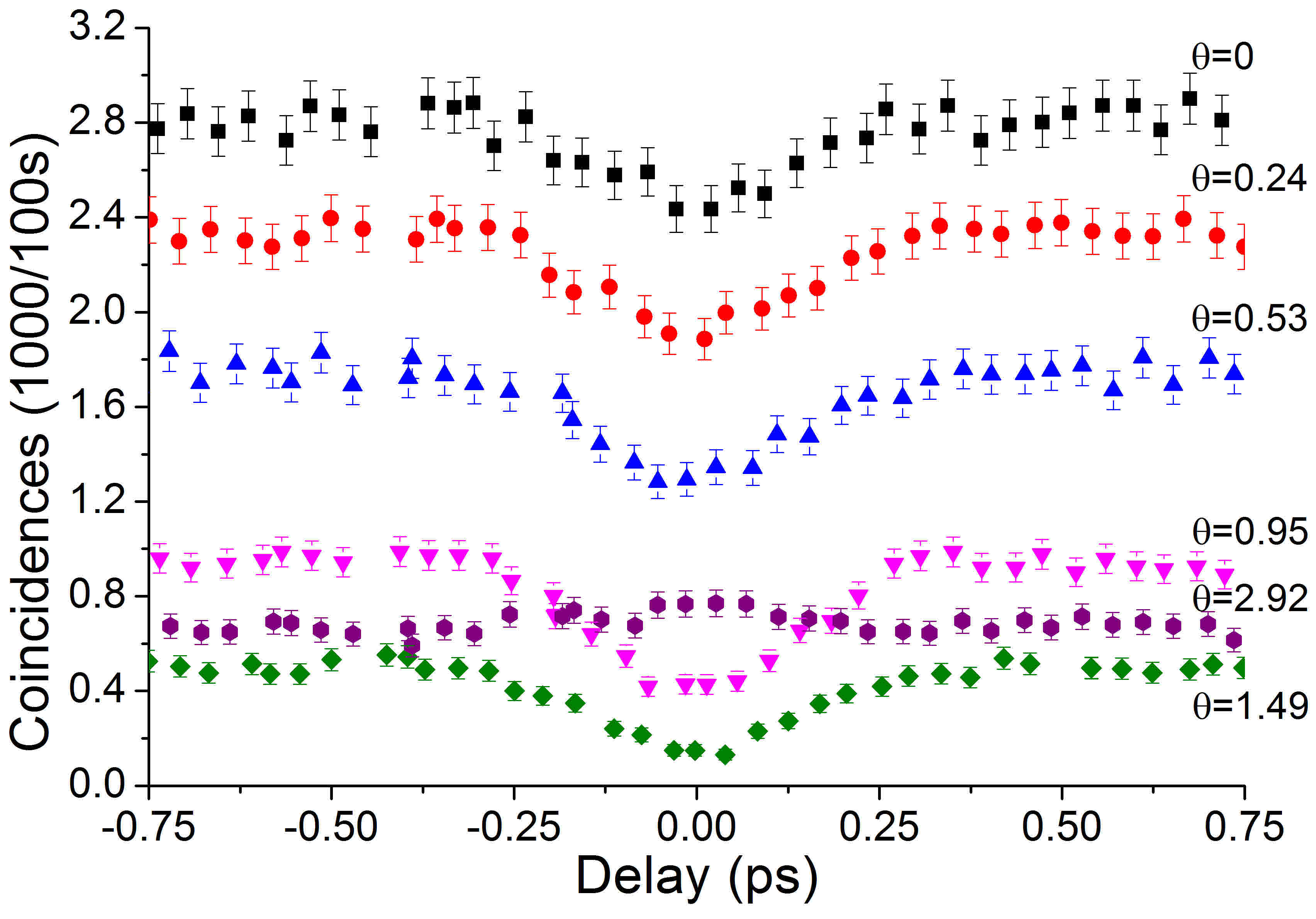}}
%  \vspace{-.4cm}
\caption{Two-photon coincidences as a function of relative delay at various values of heater voltage while injecting $\left| 011\right\rangle$ and (a) measuring $\left| 110\right\rangle$, (b) measuring $\left| 011\right\rangle$, (c) measuring $\left| 101\right\rangle$.
\label{fig:HOMDipsin23}
}
\end{figure*}

\begin{figure*}[htbp]
  \centering
  \subfigure[]{\label{fig:18}\includegraphics[width=0.3\textwidth]{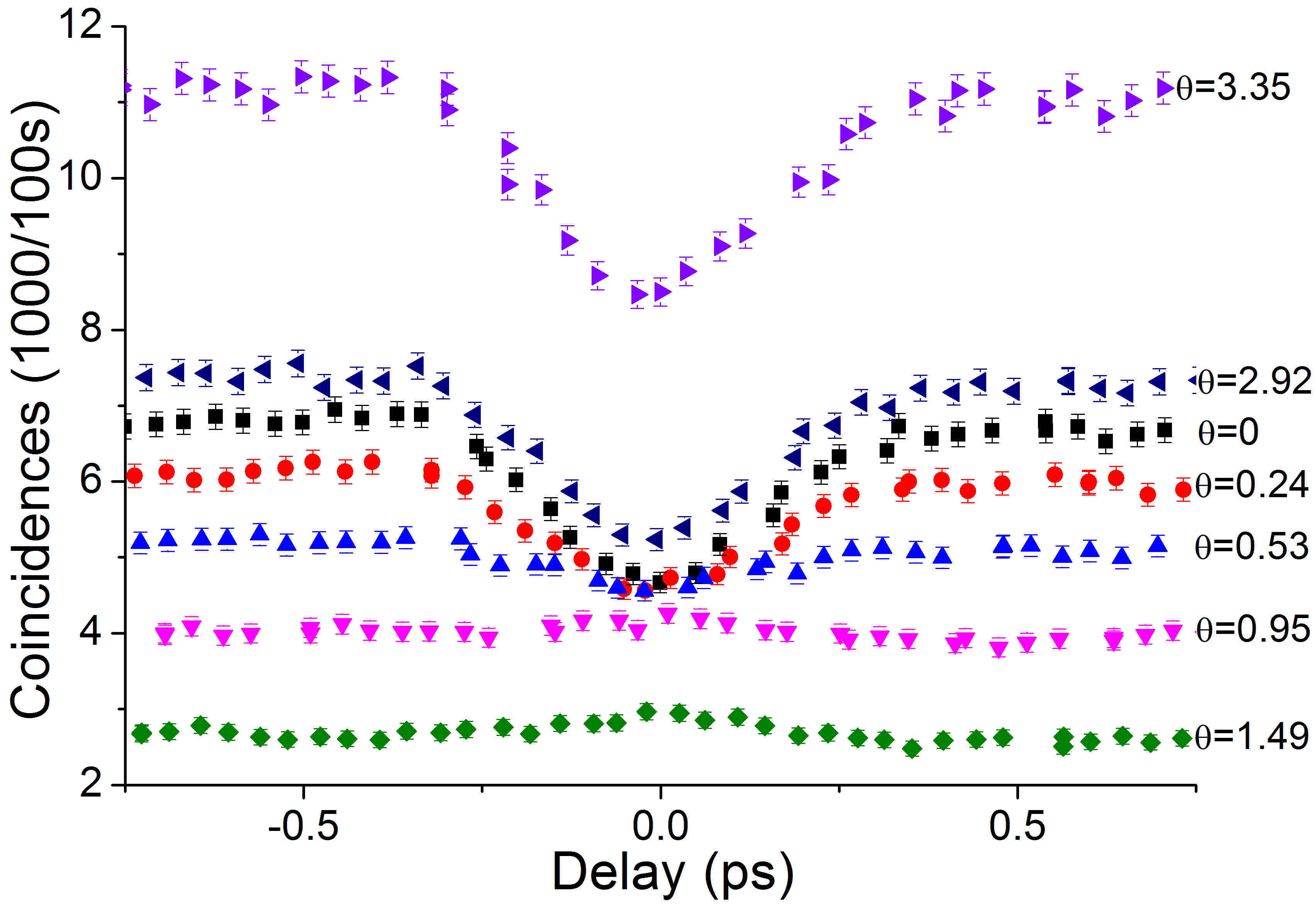}}
  \subfigure[]{\label{fig:19}\includegraphics[width=0.3\textwidth]{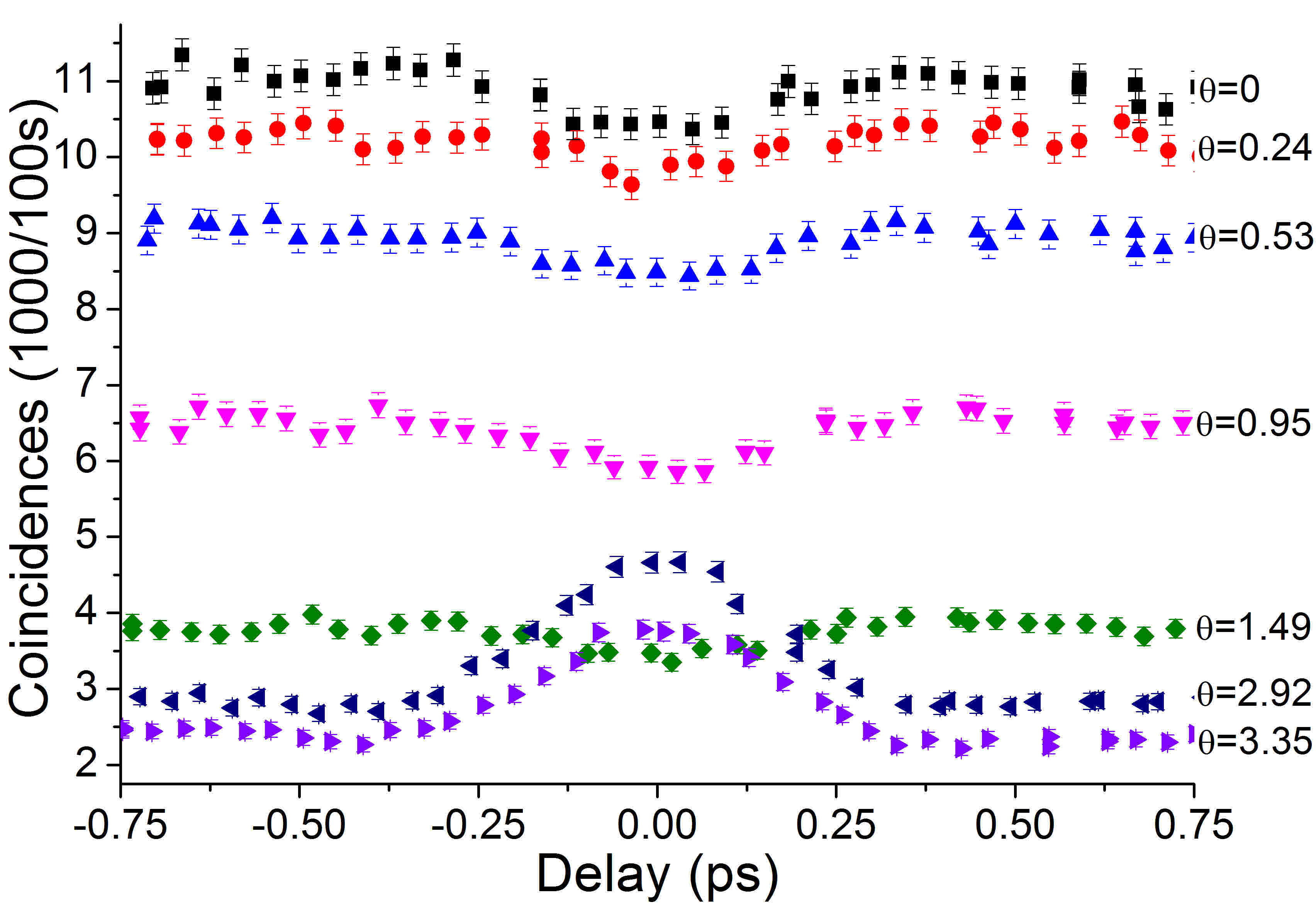}}
  \subfigure[]{\label{fig:20}\includegraphics[width=0.3\textwidth]{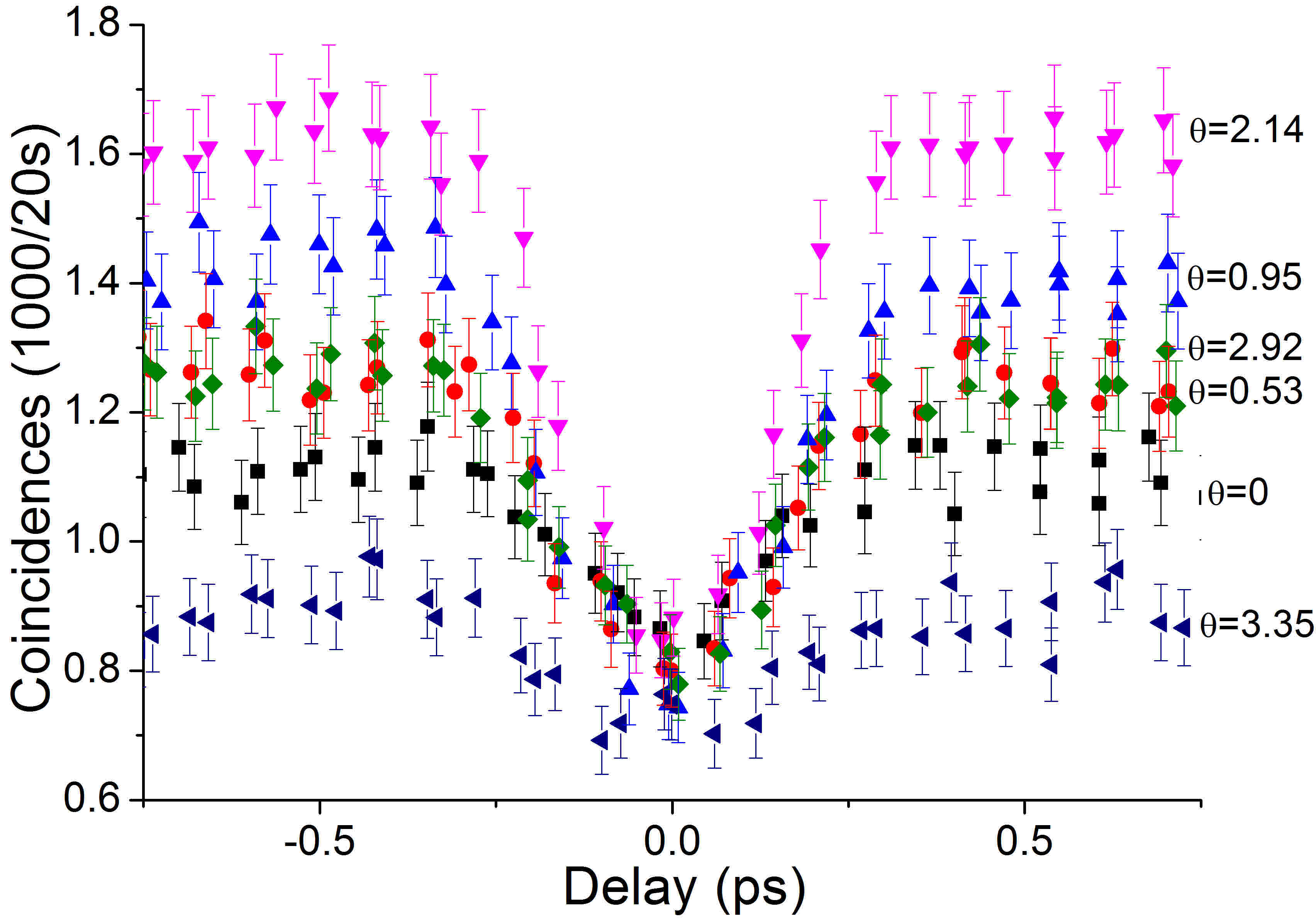}}
%  \vspace{-.4cm}
\caption{Two-fold coincidences as a function of relative delay at various values of heater voltage while injecting $\left| 101\right\rangle$ and (a) measuring $\left| 110\right\rangle$, (b) measuring $\left| 011\right\rangle$, (c) measuring $\left| 101\right\rangle$.
\label{fig:HOMDipsin13}
}
\end{figure*}

\begin{figure*}[htbp]
  \centering
  \subfigure[]{\label{fig:21}\includegraphics[width=0.3\textwidth]{./figures/V12mn.jpg}}
  \subfigure[]{\label{fig:22}\includegraphics[width=0.3\textwidth]{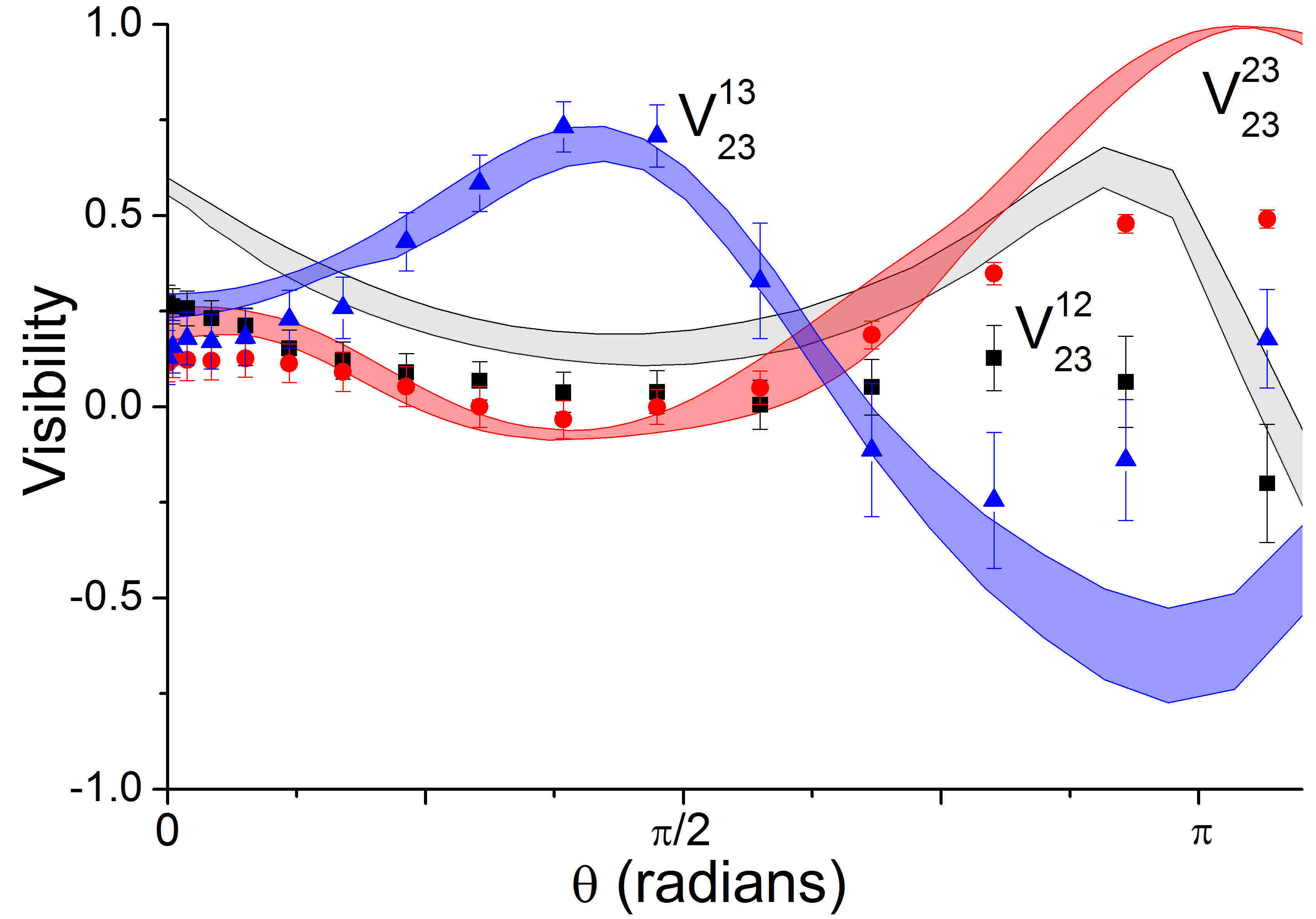}}
  \subfigure[]{\label{fig:23}\includegraphics[width=0.3\textwidth]{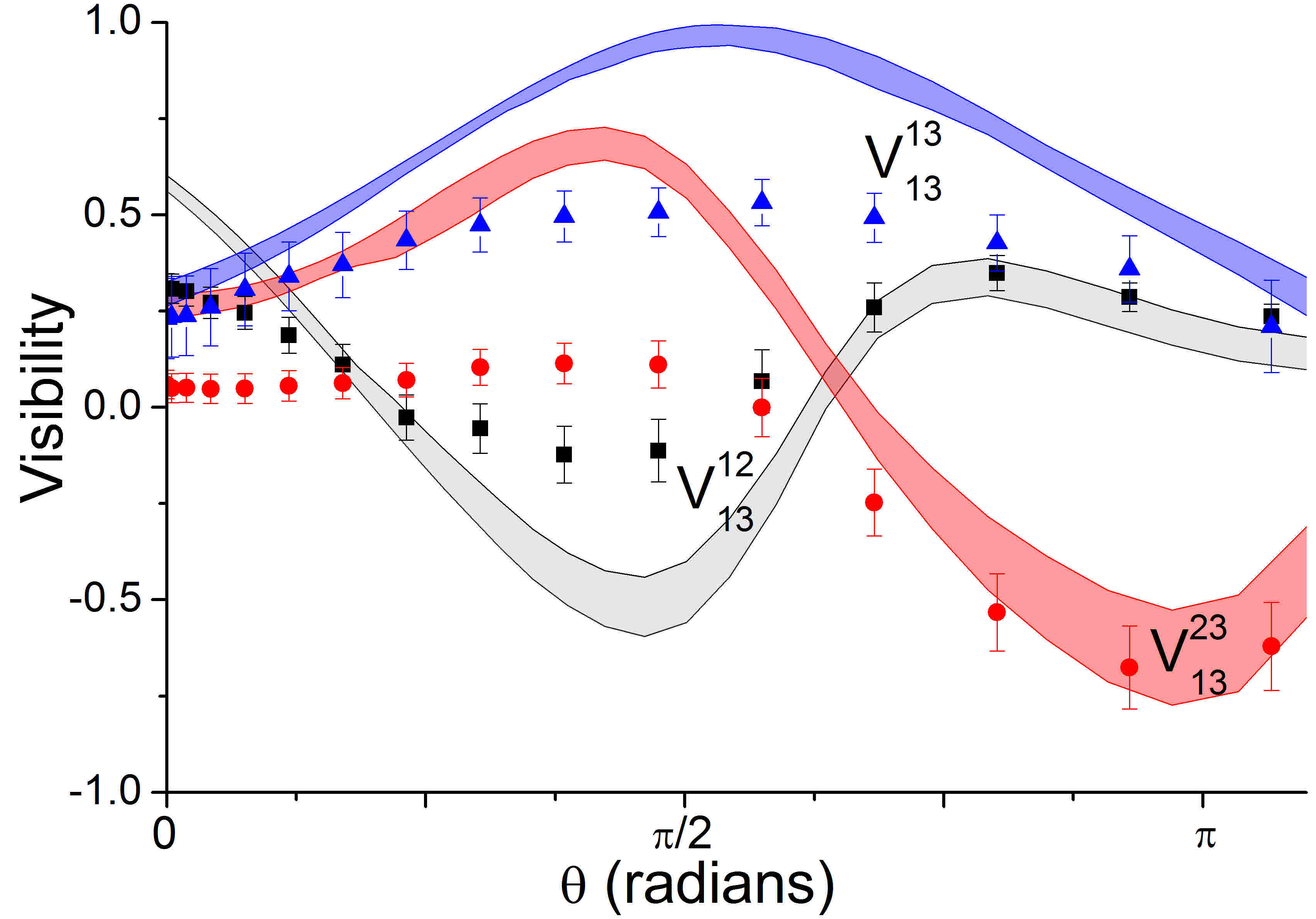}}
%  \vspace{-.4cm}
\caption{Visibilities as a function of induced phase when (a) injecting $\left| 110\right\rangle$, (b) injecting $\left| 011\right\rangle$, (c) injecting $\left| 101\right\rangle$. Black: measuring $\left| 110\right\rangle$, red: measuring $\left| 011\right\rangle$, blue: measuring $\left| 101\right\rangle$. Points: measured values determined from a Gaussian fit. Curves: predictions based on the unitary extracted from the classical characterisation.
\label{fig:Visibilities}
}
\end{figure*}

A quantum characterisation of the device was performed by injecting 804 nm photon pairs into each combination of input modes while measuring coincidences at each combination of outputs. We control the distinguishability of the photons by means of a temporal delay varied using a servo motor that translates the free-space to fibre coupling mount. A full scan of the temporal delay between the photons was performed at each value of the induced phase. Control of the two-photon interference by means of the phase shifter is evidenced by the observed change in the visibility and the transition from interference dips to peaks (Figs.~\ref{fig:HOMDipsin12}-\ref{fig:HOMDipsin13}). The visibility of each two-photon scan was determined by fitting each set of coincidence counts to a Gaussian and then comparing the counts at maximal delay $C_{mn}(\tau_\text{max})$ (the baseline) to those at minimal delay $C_{mn}(\tau=0)$ (corresponding to the extrema of the Gaussian fit)

\begin{equation}
V_{ij}^{mn}=\dfrac{C_{mn}(\tau_\text{max})-C_{mn}(0)}{C_{mn}(\tau_\text{max})}
\end{equation}
where a negative value corresponds to a coincidence peak and a positive value denotes a coincidence dip. The transition from the coalescence effect of reduced coincidences to an enhancement of coincidences with the induced phase can be understood mathematically by considering a two-photon state being injected into an arbitrary optical multiport network. When two otherwise indistinguishable photons are incident at inputs $i$ and $j$ of a network described by a unitary matrix $U$, we find that the probability of finding photons at output ports $m$ and $n$ is
\begin{equation}
p_{ij}^{mn}(\tau)=|U_{im}U_{jn}|^{2}+|U_{in}U_{jm}|^{2}+2\mathrm{Re}(U_{im}U_{jn}U_{jm}^{*}U_{in}^{*})e^{-\kappa\tau^{2}/2}
\end{equation}
where $\kappa$ is a constant related to the frequency distribution of the photons. We see that this approaches the classical value as the delay becomes infinitely large, while the quantum interference term (third) becomes important as the delay approaches zero, corresponding to the case of minimum distinguishability. It can also be seen that the interference term can take a positive or negative value depending on the complex phase dependence of the matrix elements $U_{ij}$, allowing for either destructive interference resulting in a coincidence dip or constructive interference leading to a coincidence peak. The visibility of the measured quantum interference can be predicted using the extracted unitary according to the definition above
\begin{equation}
\begin{split}
V_{ij}^{mn}&=\dfrac{p_{ij}^{mn}(\tau_\text{max})-p_{ij}^{mn}(\tau=0)}{p_{ij}^{mn}(\tau_\text{max})}\\
           &=-\dfrac{2\mathrm{Re}(U_{im}U_{jn}U_{jm}^{*}U_{in}^{*})}{|U_{im}U_{jn}|^{2}+|U_{in}U_{jm}|^{2}}
\end{split}
\end{equation}
The predicted and measured visibilities for each combination of input and output ports are plotted in Fig.~\ref{fig:Visibilities}.

\section{Extracted Fisher Information}
\begin{figure}
  \centering
  \includegraphics[width=0.4\textwidth]{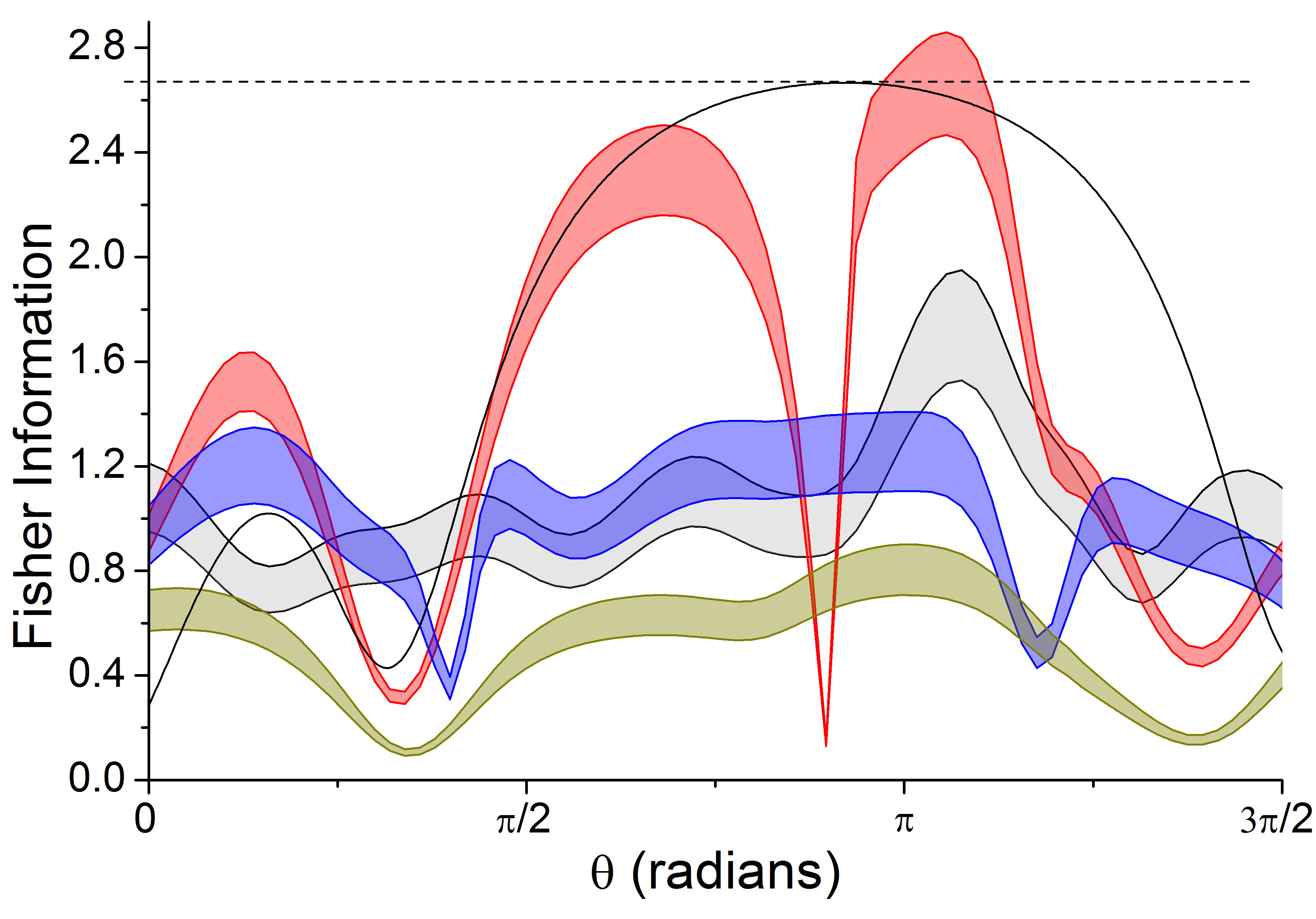}
\caption{Fisher information for two-photon input states calculated from the extracted unitary. Grey band: injecting $\left| 110\right\rangle$, red band: injecting $\left| 011\right\rangle$, blue band: injecting $\left| 101\right\rangle$, yellow band: single photon input, black curve: calculated for $\left| 110\right\rangle$ injected into an ideal device.
\label{fig:TwoPhotonFisher}
}
\end{figure}

The Fisher information quantifies the information that can be gained about a measurand $\theta$ from a given probe system by sampling its possible measurement outcomes $x$. It is obtained from the probabilities $p(x|\theta)$ of measuring $x$ given a particular value of $\theta$ through a summation over all the possible outcomes
\begin{equation}
\label{eq:Fisher}
F(\theta)=\sum\limits_{j}\dfrac{1}{p(x_{j}|\theta)}\left( \dfrac{dp(x_{j}|\theta)}{d\theta}\right)^2 
\end{equation}
In our case, the probe system is the three-arm interferometer, the measurand is an unknown phase and the measurement outcomes are the number of photons present in each output mode. We may predict the Fisher information achievable with our device by summing over all possible sets of nonclassical interference fringes $\left\langle \psi_\text{out}\right| U(\theta)\left| \psi_\text{in}\right\rangle$. These are calculated taking the input state obtained by applying creation operators $\hat{a}_{i}$ to the vacuum in input state $i$, and mapping these to the output operators $\hat{b}_{i}$ according to the unitary transformation $U$ determined above~\cite{Brougham2010}
\begin{equation}
%\begin{split}
\hat{a}_{i}\hat{a}_{j}\left| 0 \right\rangle \rightarrow \hat{b}_{i}\hat{b}_{j}\left| 0 \right\rangle = (\sum\limits_{m=1}^{3}U_{im}\hat{a}_{m})(\sum\limits_{n=1}^{3}U_{jn}\hat{a}_{n})\left| 0 \right\rangle
%\end{split}
\end{equation}
After expanding and simplifying, we take the inner product with a given output state $\left\langle \psi_\text{out}\right| = \left\langle 0 \right|\hat{a}_{m}^{\dagger}\hat{a}_{n}^{\dagger}$. Since the terms with $m,n \neq i,j$ disappear, the probability is given by the coefficient of each term $\hat{a}_{i}\hat{a}_{j}$. The probability $p_{ij}^{mn}$ of finding photons at outputs $m$ and $n$ when exciting distinct inputs $i$ and $j$ now takes the form
\begin{equation}
\label{Ps}
p_{ij}^{mn}=\dfrac{1}{1+\delta_{mn}}|U_{im}U_{jn}+U_{in}U_{jm}|^{2}.
\end{equation}
In the case of $m \neq n$ \eqref{Ps} gives the probability of measuring a coincidence at outputs $m$ and $n$, while $m=n$ corresponds to two photons emerging from output $m$. Since the transformation $U$ was not determined in a closed form, each nonclassical fringe must again be calculated ``point-by-point'', first finding $U(\theta)=e^{-i\tilde{C}(\theta)}$ at each value of $\theta$. The Fisher information is then calculated according to~\eqref{eq:Fisher}, with the derivatives of each curve being calculated numerically. The calculated Fisher information for each possible two-photon input state $\left| 110\right\rangle$, $\left| 011\right\rangle$ and $\left| 101\right\rangle$ is shown in Fig.~\ref{fig:TwoPhotonFisher}.

\end{document}